\documentclass[aps,prb]{revtex4}

\usepackage{amsmath}
\usepackage{psfig}
\usepackage{fancyhdr}
\usepackage{rotate}
\usepackage{rotating}
\usepackage{dcolumn}
\usepackage{bm}
\usepackage{subfigure}
\usepackage{epsfig}
\usepackage{natbib}
\usepackage{graphicx}
\usepackage{color}
\usepackage{latexsym}
\usepackage{amsmath}

\newcommand{\vect}[1]{\mbox{\boldmath$#1$\unboldmath}}

\newcommand{\M}{mol~l$^{-1}$}

\newcommand{\degc}{$^\circ$C}

\newcommand{\ie}{{\it i.e.}}
\newcommand{\eg}{{\it e.g.}}

\newcommand{\bsl}{{\ell}}

\newcommand{\kb}{k}
\newcommand{\ka}{\kappa a}

\newcommand{\grad}{\vect{\nabla}}
\newcommand{\lapl}{\nabla^2}
\newcommand{\dive}{\vect{\nabla} \cdot}

\begin{document}

\preprint{Manuscript submitted to {\em J. Chem. Phys.}}

\title{Transport in polymer-gel composites: Response to a bulk
concentration gradient}

\author{Reghan J. Hill}

\affiliation{Department of Chemical Engineering and McGill Institute
  for Advanced Materials, McGill University, Montreal, Quebec, H3A 2B2,
  CANADA}

\begin{abstract}
This paper examines the response of electrolyte-saturated polymer
gels, embedded with charged spherical inclusions, to a weak gradient
of electrolyte concentration.  These composites present a model
system to study microscale electrokinetic transport processes, and a
rigorous theoretical prediction of the bulk properties will benefit
novel diagnostic applications. An electrokinetic model was presented
in an earlier publication, and the response of homogeneous
composites to a weak electric field was calculated. In this work,
the influence of the inclusions on bulk ion fluxes and the strength
of an electric field (or membrane diffusion potential) induced by
the bulk electrolyte concentration gradient are computed. Effective
ion diffusion coefficients are significantly altered by the
inclusions, so---depending on the inclusion surface charge or
$\zeta$-potential---asymmetric electrolytes can behave as
symmetrical electrolytes, and vice versa. The theory also quantifies
the strength of flow driven by concentration-gradient-induced
perturbations to the equilibrium diffuse double layers. Similarly to
diffusiophoresis, the flow may be either up or down the applied
concentration gradient.
\end{abstract}

\maketitle

\section{Introduction} \label{sec:introduction}

Permeable membranes are the basis of molecular separation and sorting
technologies, including ion exchange, gel-electrophoresis
chromatography, and dialysis~\citep{Lakshminarayanaiah:1969}. Ion
permeable membranes are also vital to batteries, fuel cells,
electrochemical sensors, and biological
cells~\citep{Sten-Knudsen:2002}. In this work, a theoretical model of
electrolyte transport in membranes comprised of a continuous polymer
gel with charged spherical inclusions is used to investigate the
influence of the inclusions on bulk ion fluxes. Because several
important characteristics of the microstructure can, in principle, be
carefully controlled, these materials present a model system for
studying fundamental aspects of electrokinetic transport processes. An
ability to tailor the microstructure {\em and} accurately predict the
bulk properties could also make these materials useful in novel
diagnostic applications~\citep[\eg,][]{Hagedorn:2005}.

Electrolyte diffusion in the absence of electromigration is
accompanied by a net flux of charge when anion and cation mobilities
are not equal. Therefore, an electric field is necessary to compensate
for the net diffusive flux of charge across a membrane. In biological
cells, ion mobilities are regulated by transmembrane proteins (ion
channels), and electrical signaling is achieved by controlling the
relative ion mobilities~\citep{Sten-Knudsen:2002}. In diagnostic
electrochemical cells, diffusion potentials are deleterious, and their
influence is usually attenuated by a salt bridge containing a
concentrated symmetrical electrolyte (\eg, KCl). When charged
inclusions are immobilized in an electrically neutral polymer gel, the
effective ion mobilities are altered, so even symmetrical electrolytes
my behave as highly asymmetric electrolytes; this paper provides a
first step toward quantifying this influence.

When aqueous NaCl, for example, is brought into contact with
negatively charged inclusions (\eg, silica), the counter-ion (Na$^+$)
is concentrated in the diffuse double layers. A gradient of
electrolyte concentration perturbs the equilibrium state, inducing
diffusion {\em and} electromigration to restore local equilibrium. At
the surfaces of inclusions facing up the concentration gradient, the
equilibrium gradient of electrostatic potential attracts counter-ions
and repels co-ions, leading to inner and outer layers of perturbed
charge density. The accompanying electric field acts on the underlying
equilibrium charge, driving layers of backward and forward
electroosmotic flow. In this work, non-linear perturbations are
neglected because the perturbing `forces' are weak compared to those
of the underlying equilibrium state. Accordingly, the theory is
limited to weak electrolyte concentration gradients.

Perturbations enhance the net flux of counter-ions (Na$^+$), because
counter-ions are accumulated (depleted) at the surfaces facing up
(down) the bulk concentration gradient. The resulting (tangential)
concentration gradient increases the net (diffusive) counter-ion
flux. Note that the accompanying perturbed charge density produces an
electric field that induces an electromigrative flux of counter-ions
down the bulk concentration gradient. Therefore, because Na$^+$ ions
have a lower mobility than Cl$^-$, both contributions to the net flux
counteract the tendency of Na$^+$ to otherwise diffuse more slowly
than Cl$^-$. Overall, negatively charged inclusions increase the
(effective) symmetry of the electrolyte. Since the degree of
electrolyte symmetry bears directly on the membrane diffusion
potential, the diffusion potential characterizes the physicochemical
state of the inclusion-electrolyte interface.

The theoretical model adopted in this work was developed and applied
in an earlier publication to examine ion fluxes and electroosmotic
flow driven by an applied electric field~\citep{Hill:2005b}. In
homogeneous membranes, diffusive and electromigrative transport are
significantly influenced by the charge of the inclusions, but not by
the hydrodynamic permeability of the polymer gel. On the other hand,
electroosmotic flow is very sensitive to the gel permeability {\em
and} reorganization of charge by diffusion and electromigration. These
characteristics are expected to prevail with the application of a bulk
concentration gradient, as examined in this paper.

Note that there is a close connection of this work to earlier studies
of {\em
diffusiophoresis}~\citep{Prieve:1984,Prieve:1987,Anderson:1989}, which
address the motion of colloidal particles induced by solute
concentration gradients. Fluid motion induced by solute concentration
gradients at stationary charged interfaces is referred to as {\em
diffusioosmosis}~\citep{Dukhin:1974}, and this, perhaps, best
characterizes the phenomena investigated here. Note that Wei and
Keh~\citep{Wei:2003} derived semi-analytical predictions of the flow
induced by solute concentration gradients in (ordered) fibrous porous
media. Their calculations are restricted to low surface potentials,
however, and their theory adopts a so-called cell model to handle
fiber-fiber interactions. In the limit of vanishing fiber volume
fraction, the fibers are intervened by pure solvent, whereas in this
work the inclusions (spheres) are mediated by an uncharged Brinkman
medium.

The paper is organized as follows. The electrokinetic model and
methodology are outlined in~\S\ref{sec:theory}. Results demonstrating
the response to a weak electrolyte concentration gradient, in the
absence of a bulk electric field, are presented
in~\S\ref{sec:concalone}. These highlight the influence of negatively
charged inclusions on the effective diffusivities of the electrolyte
(NaCl) ions. Section~\ref{sec:concalone} also examines the prevailing
electroosmotic flow, which is either backward or forward, depending on
the surface charge, gel permeability, and ionic
strength. Superposition of two linearly independent problems is used
in~\S\ref{sec:diffusionandelectromigration} to solve the problem with
co-linear bulk electrolyte concentration and electrostatic potential
gradients. With the constraint of zero bulk current density, the ratio
of the electric field strength to the concentration gradient is
established. A brief summary follows in~\S\ref{sec:summary}.

\section{Theory} \label{sec:theory}

\begin{figure}
  \begin{center}
    \input{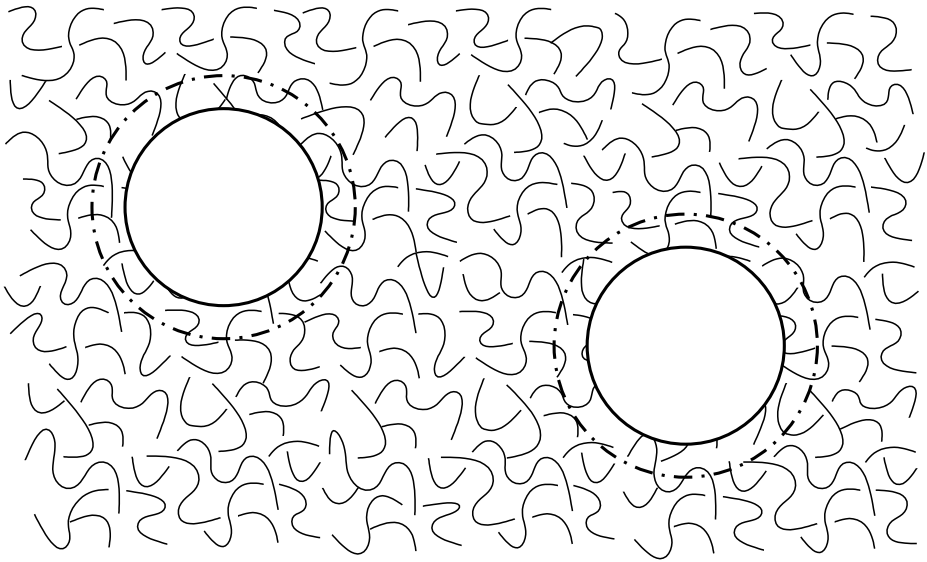}
    \caption{\label{fig:figure1} Schematic of the {\em microscale}
      under consideration. Charged, impenetrable inclusions ({\em
      solid circles}) with radius $a \sim 10$~nm--$10~\mu$m are
      embedded in a continuous polymer gel ({\em solid filaments})
      saturated with aqueous electrolyte. Diffuse double layers ({\em
      dash-dotted circles}) with thickness $\kappa^{-1} \sim
      1$--100~nm are perturbed by the application of an average
      electric field $-\langle \grad \psi \rangle$, pressure gradient
      $\langle \grad p \rangle$, or electrolyte concentration gradient
      $\langle \grad n_j \rangle$. The Brinkman screening length $\bsl
      \sim 1$--10~nm that specifies the Darcy permeability $\bsl^2$ of
      the gel is often small compared to the inclusion radius.}
  \end{center}
\end{figure}

The microstructure of the composites addressed in this work is
depicted schematically in figure~\ref{fig:figure1}. The continuous
phase is a porous medium comprised of an electrically neutral,
electrolyte-saturated polymer gel. Polyacrylamide gels are routinely
used for the electrophoretic separation of DNA fragments in aqueous
media. The porosity is controlled by adjusting the concentrations and
ratio of the monomer (acrylamide) and cross-linker. In this work, the
hydrodynamic permeability is characterized by the Darcy permeability
$\bsl^2$ (square of the Brinkman screening length), which reflects the
hydrodynamic size $a_s$ and concentration $n_s$ of the polymer
segments. In turn, these reflect the degree of cross-linking and the
affinity of the polymer for the solvent.

Embedded in the polymer are randomly dispersed spherical
inclusions. In model systems, these inclusions are envisioned to be
monodisperse silica beads or polymer latices, with radii in the range
$a = 10$~nm--$10~\mu$m. The inclusions bear a surface charge when
dispersed in aqueous media, and the surface charge density may vary
with the bulk ionic strength and pH of the electrolyte. In this work,
the charge is to be inferred from the bulk ionic strength and
electrostatic surface potential, $\zeta$. Because the inclusions are
impenetrable with zero surface capacitance and conductivity, no-flux
and no-slip boundary conditions apply at their surfaces.

Note that the mobile ions whose charge is opposite to the
surface-bound immobile charge are referred to as {\em counter-ions},
with the other species referred to as {\em co-ions}. For simplicity,
the counter-charges (dissociated counter-ions) are assumed
indistinguishable from the electrolyte counter-ions. Surrounding each
inclusion is a diffuse layer of mobile charge, with Debye thickness
$\kappa^{-1}$ and excess of counter-ions; as described below, the
equilibrium double-layer structure is calculated from the well-known
Poisson-Boltzmann equation.

\subsection{Microscale model}

The microscale transport equations and boundary conditions are based
on the {\em standard electrokinetic
model}~\citep{Overbeek:1943,Booth:1950} with a body force to model the
frictional resistance of the
polymer~\citep{Brinkman:1947,Debye:1948}. This coupling has been
widely used to interpret the electrophoretic mobility and other
characteristics of `soft' colloidal particles and their
dispersions~\citep[\eg,][]{Wunderlich:1982,Levine:1983,Ohshima:1989,Saville:2000,Hill:2003a}. The
model comprises the (non-linear) Poisson-Boltzmann equation
\begin{equation} \label{eqn:pbeqn}
  \epsilon_o \epsilon_s \lapl \psi = - \sum_{j=1}^{N} n_j z_j e,
\end{equation}
where $\epsilon_o$ and $\epsilon_s$ are the permittivity of a vacuum
and dielectric constant of the electrolyte; $n_j$ are the
concentrations of the $j$th mobile ions with valences $z_j$; and
$\psi$ and $e$ are the electrostatic potential and elementary
charge. Ion transport is governed by the Nernst-Plank relationship
\begin{equation} \label{eqn:iontransport}
  6 \pi \eta a_j (\vect{u} - \vect{v}_j) - z_j e \grad \psi - \kb T
  \grad \ln{n_j} = 0 \ \ (j=1,...,N),
\end{equation}
where $a_j$ are the Stokes radii of the ions, obtained from limiting
conductances or diffusivities; $\eta$ is the electrolyte viscosity;
$\vect{u}$ and $\vect{v}_j$ are the fluid and ion velocities; and $\kb
T$ is the thermal energy. Ion diffusion coefficients, which will be
adopted throughout, are
\begin{equation} \label{eqn:diffusivity}
  D_j = \kb T / (6 \pi \eta a_j).
\end{equation}

As usual, the double-layer thickness (Debye length)
\begin{equation} \label{eqn:kappa}
  \kappa^{-1} = \sqrt{\kb T \epsilon_s \epsilon_o / (2 I e^2)}
\end{equation}
emerges from Eqns.~(\ref{eqn:pbeqn}) and~(\ref{eqn:iontransport})
where
\begin{equation} \label{eqn:ionicstrength}
  I = (1/2) \sum_{j=1}^{N} z^2_j n^\infty_j
\end{equation}
is the bulk (average) ionic strength, with $n_j^\infty$ the bulk ion
concentrations.

Ion conservation demands
\begin{equation} \label{eqn:ionconservation}
  \partial n_j / \partial t = 0 = -\dive (\vect{j}_j) \ \ (j=1,...,N),
\end{equation}
where $t$ is the time, with the ion fluxes $\vect{j}_j = n_j
\vect{v}_j$ obtained from Eqn.~(\ref{eqn:iontransport}). Similarly,
momentum and mass conservation require
\begin{equation} \label{eqn:linnseqns}
  \rho_s \partial \vect{u} / \partial t = \vect{0} = \eta \lapl
  \vect{u} - \grad p - (\eta / \bsl^2) \vect{u} - \sum_{j=1}^{N} n_j
  z_j e \grad \psi
\end{equation}
and
\begin{equation} \label{eqn:incomp}
  \dive \vect{u} = 0,
\end{equation}
where $\rho_s$ and $\vect{u}$ are the electrolyte density and
velocity, and $p$ is the pressure. Note that $-(\eta / \bsl^2)
\vect{u}$ is the hydrodynamic drag force exerted by the polymer on the
electrolyte. The Darcy permeability (square of the Brinkman screening
length) of the gel may be expressed as
\begin{equation} \label{eqn:brinkmanscreeninglength}
  \bsl^2 = 1 / [n_s(r) 6 \pi a_s F_s] = 2 a_s^2 / [9 \phi_s(r)
    F_s(\phi_s)],
\end{equation}
where $n_s(r)$ is the concentration of Stokes resistance centers, with
$a_s$ and $F_s$ the Stokes radius and drag coefficient of the polymer
segments. The hydrodynamic volume fraction $\phi_s = n_s (4/3) \pi
a_s^3$ of the segments in swollen polymer gels is often very low, so
$F_s \approx 1$. In this work, the Brinkman screening length is
adjusted according to Eqn.~(\ref{eqn:brinkmanscreeninglength}) by
varying the (uniform) polymer segment density with Stokes radius $a_s
= 1$~\AA. The drag coefficient $F_s$ is obtained from a correlation
for random fixed beds of spheres~\citep{Koch:1999}. For the purposes
of this paper, however, only the reported values of $\bsl$ are
relevant~\citep[see][]{Hill:2004a,Hill:2005a}.

Either the equilibrium surface potential $\zeta$ or surface charge
density $\sigma$ may be specified. Because the surface ($r = a$) is
assumed impenetrable with zero capacitance and conductivity, the
surface charge is constant, permitting no-flux boundary conditions for
each (mobile) ion species. As usual, the no-slip boundary condition
applies.

In the far field, neglect of particle interactions requires
\begin{equation} \label{eqn:psibc2}
  \psi \rightarrow - \vect{E} \cdot \vect{r} \mbox{ as } r \rightarrow
  \infty,
\end{equation}
\begin{equation} \label{eqn:averageconc}
  n_j \rightarrow n_j^\infty + \vect{B}_j \cdot \vect{r} \mbox{ as } r
  \rightarrow \infty,
\end{equation}
and
\begin{equation} \label{eqn:rest}
  \vect{u} \rightarrow \vect{U} \mbox{ as } r \rightarrow \infty,
\end{equation}
where $\vect{E}$, $\vect{B}_j$ and $\vect{U}$ are, respectively, a
constant electric field, constant species concentration gradients, and
constant far-field velocity.

With `forcing'
\begin{equation}
  \vect{X} = X \vect{e}_z,
\end{equation}
where $X \in \{E, B_j, U\}$, linearized perturbations $-\vect{E} \cdot
\vect{r} + \psi'$, $\vect{B}_j \cdot \vect{r} + n_j'$ and $\vect{U} +
\vect{u}'$ to the equilibrium state ($\psi^0$, $n_j^0$ and $\vect{u} =
\vect{0}$) are symmetric about the $z$-axis ($\theta = 0$) of a
spherical polar coordinate system. Primed quantities have the form
\begin{equation} \label{eqn:pert1}
  \psi' = \hat{\psi}(r) \vect{X} \cdot \vect{e}_r
\end{equation}
\begin{equation}  \label{eqn:pert2}
  n_j' = \hat{n}_j(r) \vect{X} \cdot \vect{e}_r
\end{equation}
and
\begin{eqnarray}\label{eqn:velocity}
  \vect{u}' &=& \grad \times \grad \times h(r) \vect{X} \nonumber \\
  &=& - 2 (h_r / r) (\vect{X} \cdot \vect{e}_r) \vect{e}_r - (h_{rr} +
  h_{r} / r) (\vect{X} \cdot \vect{e}_\theta) \vect{e}_\theta.
\end{eqnarray}
The radially decaying functions $\hat{\psi}(r)$, $\hat{n}_j(r)$ and
$h(r)$ are calculated numerically in this work. However, as discussed
below, and at length elsewhere~\citep{Hill:2005b}, only the scalar
coefficients characterizing their far-field decays (see
Eqns.~(\ref{eqn:farfield1})--(\ref{eqn:farfield3}) below) are
necessary to derive bulk properties of the composite. Therefore, the
practical purpose of the numerical procedure is to determine these
so-called {\em asymptotic coefficients}.

Solutions of Eqns.~(\ref{eqn:pbeqn}), (\ref{eqn:ionconservation}),
(\ref{eqn:linnseqns}) and (\ref{eqn:incomp}), to linear order in
perturbations to the equilibrium state, with $E$, $B_j$ and $U$ set to
arbitrary values, can be computed if
\begin{equation}
  \sum_{j=1}^N z_j B_j = 0
\end{equation}
to ensure an electrically neutral far-field. However, when $N$ species
are assembled into $M$ electroneutral groups (\eg, electrolytes or
neutral tracers), each with far-field gradient $B_k$ ($k = 1,...,M$),
it is expedient to compute solutions with only one non-zero value of
$E$, $B_k$ or $U$. Then, arbitrary solutions can be constructed by
linear superposition~\citep{OBrien:1978}.

An index $k'$ is required to identify the (electroneutral) group to
which the $j$th species under consideration is assigned. Careful
consideration of the electrolyte composition and ion valences is
required to ensure consistency. For $z$-$z$ electrolytes it is
convenient to set $B_j = B_k$, whereas for a single 1-2 electrolyte
(\eg, CaCl$_2$) is it satisfactory to set $B_j = B_{k'} / |z_j|$. For
the relatively simple situations considered in this work, the (single)
electroneutral group is NaCl, so $M = 1$ with $k = k' = 1$, and $j =
1$ and 2 for Na$^+$ and Cl$^-$, respectively.

The perturbations satisfy a linear set of coupled ordinary
differential equations with far-field boundary
conditions~\citep{Hill:2005b}
\begin{equation} \label{eqn:farfield1}
  \psi' \rightarrow (\vect{X} \cdot \vect{e}_r) D^X / r^2 \mbox{ as }
  r \rightarrow \infty,
\end{equation}
\begin{equation} \label{eqn:farfield2}
  n'_j \rightarrow (\vect{X} \cdot \vect{e}_r) C^X_j / r^2 \mbox{ as }
  r \rightarrow \infty,
\end{equation}
and
\begin{equation}  \label{eqn:farfield3}
  \vect{u}' \rightarrow - 2 (C^X / r^3) (\vect{X} \cdot \vect{e}_r)
  \vect{e}_r + (C^X/r^3) (\vect{X} \cdot \vect{e}_\theta)
  \vect{e}_\theta \mbox{ as } r \rightarrow \infty.
\end{equation}
The equations for the non-linear equilibrium state and the linearized
perturbations are solved using the numerical methodology of Hill,
Saville and Russel~\cite{Hill:2003a}, which was developed for the
electrophoretic mobility of polymer-coated colloids and the bulk
properties of their dilute dispersions.

Note that $C_j^{B_k}$ is the asymptotic coefficient for the perturbed
{\em concentration} of the $j$th species induced by the $k$th
concentration gradient $B_k$, whereas $C^{B_k}$ (without a subscript)
denotes the asymptotic coefficient for the {\em flow} induced by
$B_k$. For neutral species, the concentration disturbance produced by
a single impenetrable sphere yields $C_j^{B_{k'}} = (1/2) a^3$,
otherwise $C_j^{B_{k}} = 0$ ($k \ne k'$). Clearly, the asymptotic
coefficients for charged species, whose concentration perturbations
are influenced by electromigration, are not the same as for neutral
species; for ions, however, $C_j^{B_{k'}} \rightarrow (1/2) a^3$ as
$|\zeta| \rightarrow 0$.

With co-linear forcing and bulk electroneutrality, linear
superposition gives far-field decays
\begin{equation}
  \psi' \rightarrow (1 / r^2)[E D^E + \sum_{k=1}^{M} B_k D^{B_k} + U
    D^U] (\vect{e}_z \cdot \vect{e}_r) \mbox{ as } r \rightarrow
  \infty,
\end{equation}
\begin{equation}
  n_j' \rightarrow (1 / r^2)[E C_j^E + \sum_{k=1}^{M} B_k C_j^{B_k} +
    U C_j^U] (\vect{e}_z \cdot \vect{e}_r) \mbox{ as } r \rightarrow
  \infty,
\end{equation}
and
\begin{eqnarray} 
  \vect{u}' \rightarrow - (2 / r^3) [E C^E + \sum_{k=1}^{M} B_k
    C^{B_k} + U C^U] (\vect{e}_z \cdot \vect{e}_r) \vect{e}_r
  \nonumber \\ + (1/r^3) [E C^E + \sum_{k=1}^{M} B_k C^{B_k} + U
    C^U] (\vect{e}_z \cdot \vect{e}_\theta) \vect{e}_\theta \mbox{ as
  } r \rightarrow \infty.
\end{eqnarray}
These relationships are the key results from which all macroscale
quantities (\eg, bulk ion fluxes and electroosmotic flow) are derived
for small inclusion volume fractions~\citep{Hill:2005b}. Situations
with only one forcing variable $X \in \{E, B_k, U\}$ are referred to
as the (E), (B) and (U) (microscale) problems. Algebraic or
differential relationships between the averaged fields can be applied
to ensure, for example, zero average current density
(see~\S~\ref{sec:diffusionandelectromigration}).

\subsection{Macroscale equations}

Expressions relating the asymptotic coefficients $D_j^X$, $C_j^X$ and
$C^X$ to the averaged ion fluxes and momentum conservation equation
can be derived using procedures similar to those applied by
Saville~\cite{Saville:1979} and O'Brien~\cite{OBrien:1981} for the
conductivity of colloidal dispersions. As demonstrated in an earlier
paper~\citep{Hill:2005b}, ion and fluid momentum fluxes can be
averaged over a representative elementary control volume, and the
volume integrals enumerated from knowledge of the asymptotic
coefficients. Similarly to Maxwell's well-known analysis of
conduction, particle interactions are neglected, so the results are
limited to small volume fractions $\phi [1 + (\ka)^{-1}]^3 \ll 1$,
where $\phi = n (4/3) \pi a^3$ is the inclusion volume fraction, with
$n$ the inclusion number density.

When {\em all} average fluxes are in the $z$-direction, mass and
momentum conservation require constant average velocity $\langle
\vect{u} \rangle$, giving~\citep{Hill:2005b}
\begin{eqnarray} \label{eqn:avemomC}
  \langle \grad{p} \rangle = - (\eta / \bsl^2) \langle \vect{u}
  \rangle - \phi (3/a^3) (\eta / \bsl^2) [\vect{E} C^E +
    \sum_{k=1}^{M} \vect{B}_k C^{B_{k}} + \vect{U} C^U],
\end{eqnarray}
where $\langle \grad{p} \rangle$ is the average pressure gradient (set
to zero in this work). Similarly, the (steady) average species
conservation equations $\dive{\langle \vect{j}_j \rangle} = 0$ require
constant average fluxes~\citep{Hill:2005b}
\begin{eqnarray}  \label{eqn:finalfluxB}
  \langle \vect{j}_j \rangle = n_j^\infty \langle \vect{u} \rangle -
  z_j e \frac{D_j}{\kb T} n_j^\infty \langle \grad{\psi} \rangle - D_j
  \langle \grad n_j \rangle \nonumber \\ + \phi (3/a^3) z_j e
  \frac{D_j}{\kb T} n_j^\infty [\vect{E} D^E + \sum_{k=1}^{M}
  \vect{B}_{k} D^{B_{k}} + \vect{U} D^U] \nonumber \\ + \phi (3 / a^3)
  D_j [\vect{E} C_j^E + \sum_{k=1}^M \vect{B}_k C_{j}^{B_{k}} +
  \vect{U} C_j^U].
\end{eqnarray}
Note that $\dive{\langle \grad \psi \rangle} = 0$ in an electrically
neutral composite with uniform dielectric permittivity, so the average
electric field $\langle \grad \psi \rangle$ is also constant.

The averages can be expanded as power series in the inclusion volume
fraction \eg, $\langle \vect{u} \rangle \rightarrow \vect{U}_0 + \phi
\vect{U}_1 + O(\phi^2)$. Therefore, since the microscale equations
(asymptotic coefficients) are accurate to $O(\phi)$, the notation is
condensed by writing, for example, $\langle \vect{u} \rangle \equiv
\vect{U}$, where it is understood that $\vect{U} = \vect{U}_0 + \phi
\vect{U}_1 + O(\phi^2)$. Clearly, $\vect{E}$, $\vect{B}_j$ and
$\vect{U}$ in Eqns.~(\ref{eqn:avemomC}) and (\ref{eqn:finalfluxB})
need only include the $O(1)$ contribution to their respective average
value, \eg, $\vect{U} \rightarrow \vect{U}_0$. The following notation
is adopted for the other averaged quantities: $\vect{J}_j \equiv
\langle \vect{j}_j \rangle$, $\vect{P} \equiv \langle \grad p
\rangle$, $\vect{B}_j \equiv \langle \grad{n}_j \rangle$, $\vect{E}
\equiv - \langle \grad \psi \rangle$.

With one electrolyte ($M=1$) and, recall, bulk electroneutrality,
there are $N + 4$ independent variables ($\vect{E}, \vect{U},
\vect{P}, \vect{B}_k~(k=1), \vect{J}_j$~($j=1,...,N$)) with $N + 1$
independent equations (Eqns.~(\ref{eqn:avemomC}) and
(\ref{eqn:finalfluxB})). Clearly, three independent variables must be
specified for a unique solution.

For clarity, the results presented below involve a 1-1 electrolyte
(NaCl). It is important to note that, because the equations are
linear, solutions for any combination of non-zero forcing variables
may be constructed. For example,
\S\ref{sec:diffusionandelectromigration} establishes the electric
field strength required to maintain a constant electrolyte
flux---driven by a bulk concentration gradient across a
membrane---with zero bulk current density. Solutions of the (E)
problem are available elsewhere~\citep{Hill:2005b}, whereas all
solutions of the (B) problem presented below are new.

\section{Results}

\subsection{Concentration gradient alone} \label{sec:concalone}

When an average concentration gradient is applied in the absence of an
average pressure gradient and electric field, the ion fluxes
(Eqn.~(\ref{eqn:finalfluxB})) are 
\begin{eqnarray} \label{eqn:bfieldflux}
  \vect{J}_j = - D_j \vect{B}_{j} + \phi (3/a^3) z_j e \frac{D_j}{\kb
    T} n_j^\infty \sum_{k=1}^{M} \vect{B}_{k} D^{B_k} \nonumber \\ +
  \phi (3 / a^3) D_j \sum_{k=1}^{M} \vect{B}_{k} C_{j}^{B_{k}} +
  \phi (3/ a^3) n_j^\infty \sum_{k=1}^{M} \vect{B}_k C^{B_k} +
  O(\phi^2).
\end{eqnarray}
The first term on the right-hand side is the diffusive flux in the
absence of inclusions, and the second, third and fourth terms,
respectively, are corrections due to microscale electromigration,
diffusion and convection due to the inclusions.

For example, with the composite bridging two reservoirs, one
containing KCl with concentrations $n_{j,1}^\infty$ at position $z =
0$, and the other containing NaCl with concentrations $n_{j,2}^\infty$
at $z = L$, it is necessary to designate $M = 2$ bulk electrolyte
concentration gradients for the $N = 3$ species: $B_1 = -
n_{j,1}^\infty / L$ for KCl, and $B_2 = n_{j,2}^\infty / L$ for
NaCl. Therefore, the bulk concentration gradients for K$^+$, Na$^+$,
and Cl$^-$ are, respectively, $B_1$, $B_2$ and $B_1 + B_2$.

For the simple case involving the concentration gradient of a single
$z$-$z$ electrolyte ($B_j = B_{k'}$, $N=2$ and $M=1$),
Eqn.~(\ref{eqn:bfieldflux}) may be written
\begin{equation}
  \vect{J}_j = - D_j \vect{B}_{j} (1 + \phi \Delta_j^{B_{k'}}),
\end{equation}
where
\begin{eqnarray} \label{eqn:bfieldincrements}
  \Delta_j^{B_{k'}} &=& \Delta_{j,e}^{B_{k'}} + \Delta_{j,d}^{B_{k'}}
  + \Delta_{j,c}^{B_{k'}} \nonumber \\ &=& - (3/a^3) \frac{z_j e
    n_j^\infty}{\kb T} D^{B_{k'}} - (3 / a^3) C_{j}^{B_{k'}} - (3/ a^3)
  \frac{n_j^\infty}{D_j} C^{B_{k'}}.
\end{eqnarray}
This motivates the introduction of an effective diffusivity
\begin{equation}
  D_j^* = - J_j / B_{j} = D_j (1 + \phi \Delta_j^D),
\end{equation}
where $\Delta_j^D = \Delta_j^{B_{k'}}$ is termed the (dimensionless)
{\em effective diffusivity increment}.

Note that electromigration from concentration-gradient-induced
electrical polarization increases (decreases) the bulk counter-ion
(co-ion) flux, leading to a (bulk) current density
\begin{eqnarray} \label{eqn:diffusioncurrent}
  \vect{I} &=& \sum_{j=1}^N z_j e \vect{J}_j \\ &=& - \vect{B}_{k'}
  \sum_{j=1}^N z_j e D_j (1 + \phi \Delta_j^{B_k'}). \label{eqn:diffusioncurrentp}
\end{eqnarray}
An average concentration gradient also induces an $O(\phi)$ average
velocity that reflects the permeability $\bsl^2$ of the polymer
gel. It is therefore expedient to examine the ratio of the average
fluid velocity to the product of the particle volume fraction and the
strength of the concentration gradient
\begin{equation} \label{eqn:mobility}
  U / (B_{k'} \phi) = -3 C^{B_{k'}} / a^3.
\end{equation}
This quantity, termed the {\em incremental pore mobility}, is similar
to the incremental pore mobility that characterizes
electric-field-induced flow~\citep{Hill:2005b}. It is also bears
similarity to the diffusiophoretic mobility of free colloidal
particles~\citep{Prieve:1987}.

Note that Eqns.~(\ref{eqn:diffusioncurrentp}) and~(\ref{eqn:mobility})
apply with zero average electric field. This implies that an electric
field is impressed to counteract the field that would otherwise
develop spontaneously to achieve zero bulk electrical current. The
pore mobility with zero electrical current, for example, may be
obtained by eliminating $\vect{E}$ from Eqns.~(\ref{eqn:avemomC})
and~(\ref{eqn:finalfluxB}), with $\vect{P} = \vect{I} = \vect{0}$, as
demonstrated in~\S\ref{sec:diffusionandelectromigration}. For the
simple case involving a single $z$-$z$ electrolyte ($B_j = B_{k'}$,
$N=2$ and $M=1$), the pore mobility is
\begin{equation} \label{eqn:mobilityzerocurrent}
  U / (B_{k'} \phi) = -(3 / a^3) [C^{B_{k'}} + C^E \sum_{j=1}^{N=2}
    z_j e D_j / K^\infty] + O(\phi),
\end{equation}
where
\begin{equation} \label{eqn:bulkcond}
  K^\infty = \sum_{j=1}^{N} (z_j e)^2 \frac{D_j}{\kb T} n_j^\infty
\end{equation}
is the electrolyte conductivity.

\begin{figure}
  \begin{center}
    \vspace{1.0cm}
    \includegraphics[width=6cm]{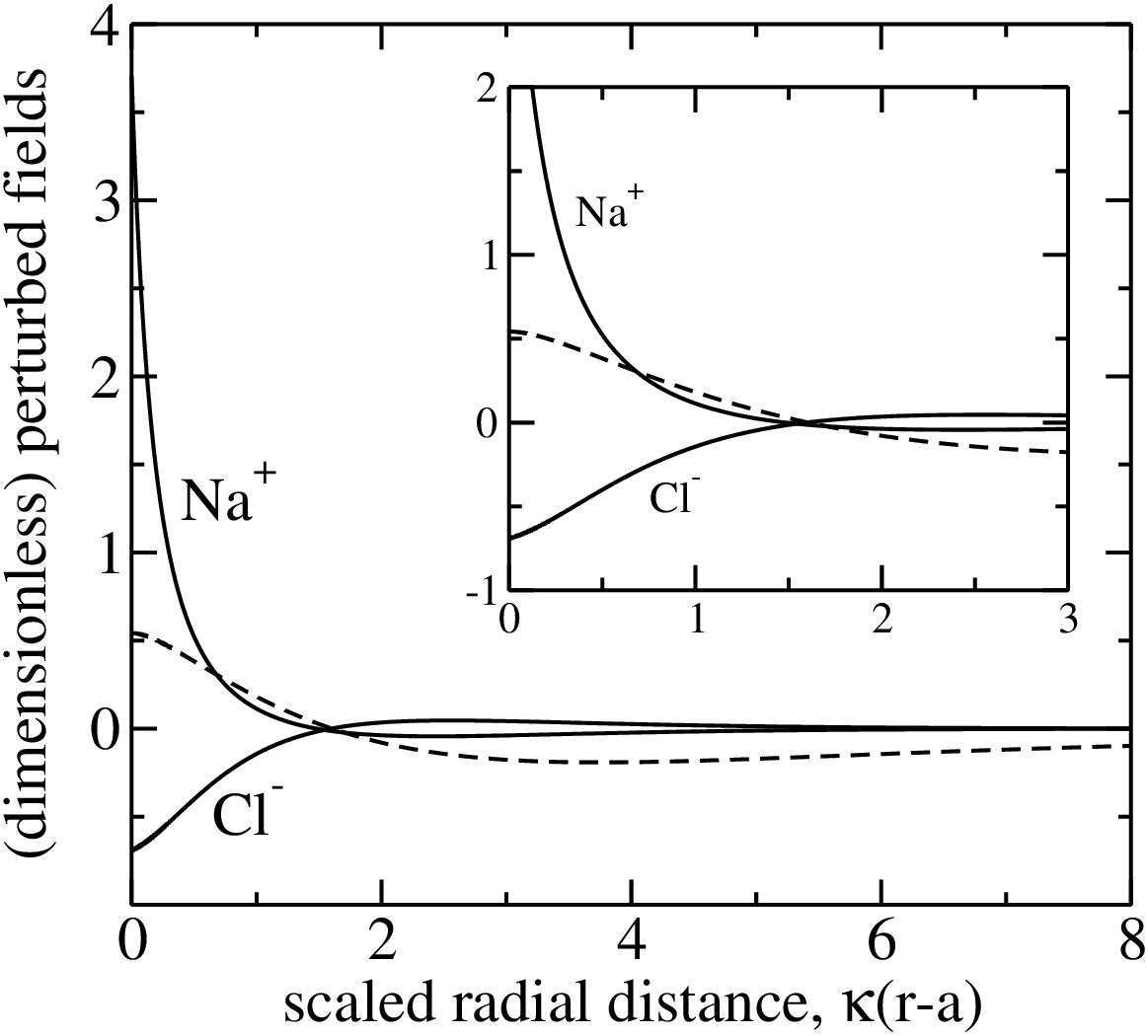} \hspace{1.0cm} \includegraphics[width=6cm]{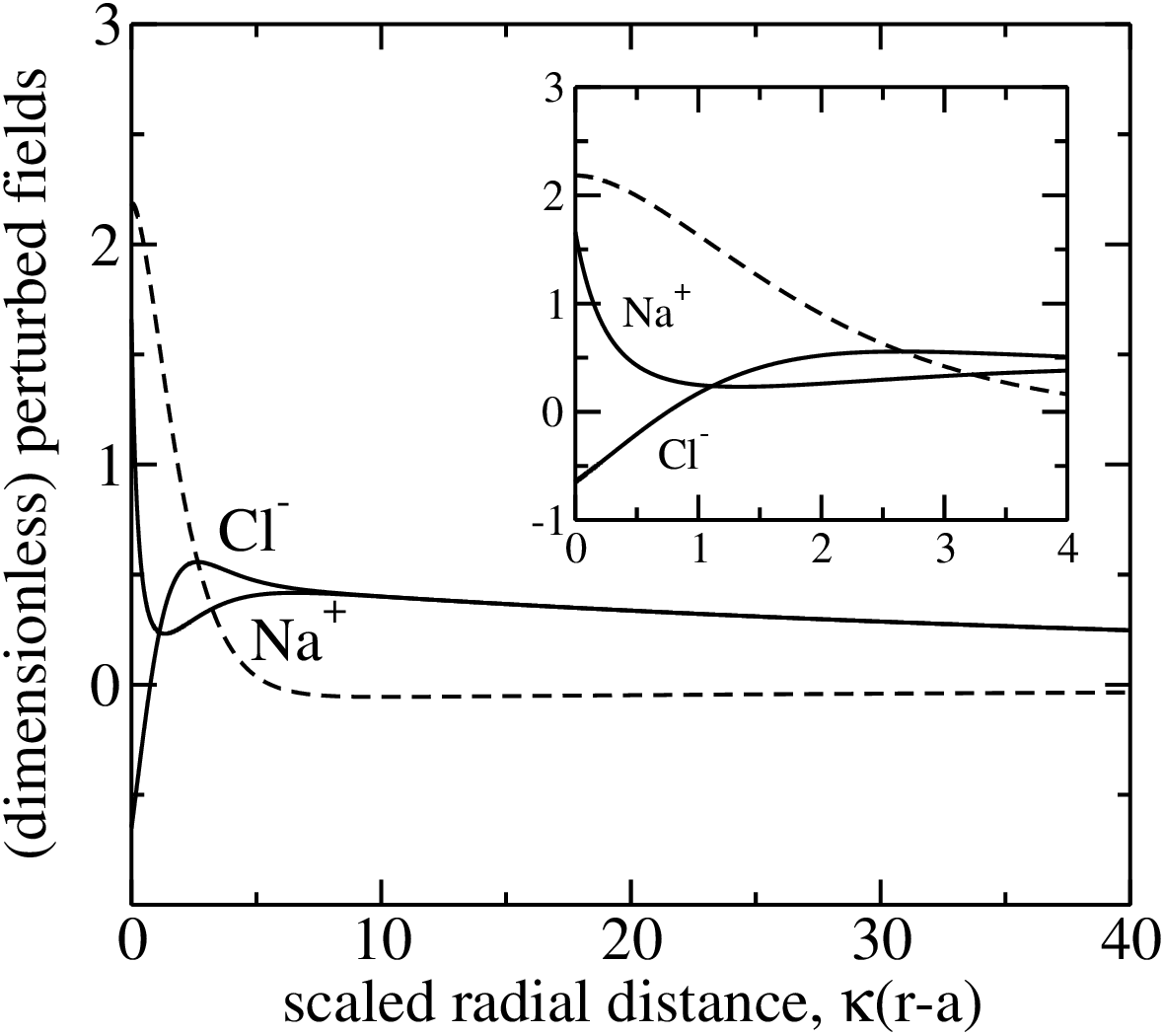}
    \vspace{0.5cm}
  \end{center}
  \caption{\label{fig:perturbeda} The structure of the perturbed
    double layer---under the influence of a bulk electrolyte
    concentration gradient---as revealed by the (scaled) radial
    electrostatic potential, $\hat{\psi}(r) 2 I e / (\kb T a)$ ({\em
    dashed lines}), and ion concentration, $\hat{n}_1(r) / a$~(Na$^+$)
    and $\hat{n}_2(r) /a$~(Cl$^+$) ({\em solid lines}), perturbations
    (see Eqns.~(\ref{eqn:pert1})--(\ref{eqn:velocity})) as a function
    of the (scaled) radial position $\kappa (r - a)$: aqueous NaCl at
    $T = 25$\degc; $a = 100$~nm; $\bsl \approx 3.11$~nm; $(\ka, \zeta
    e / (\kb T)) = (1, -2)$ ({\em left panel}); $(\ka, \zeta e / (\kb
    T)) = (100, -2)$ ({\em right panel}).}
\end{figure}

\begin{figure}
  \begin{center}
    \vspace{1.0cm}
    \includegraphics[width=6cm]{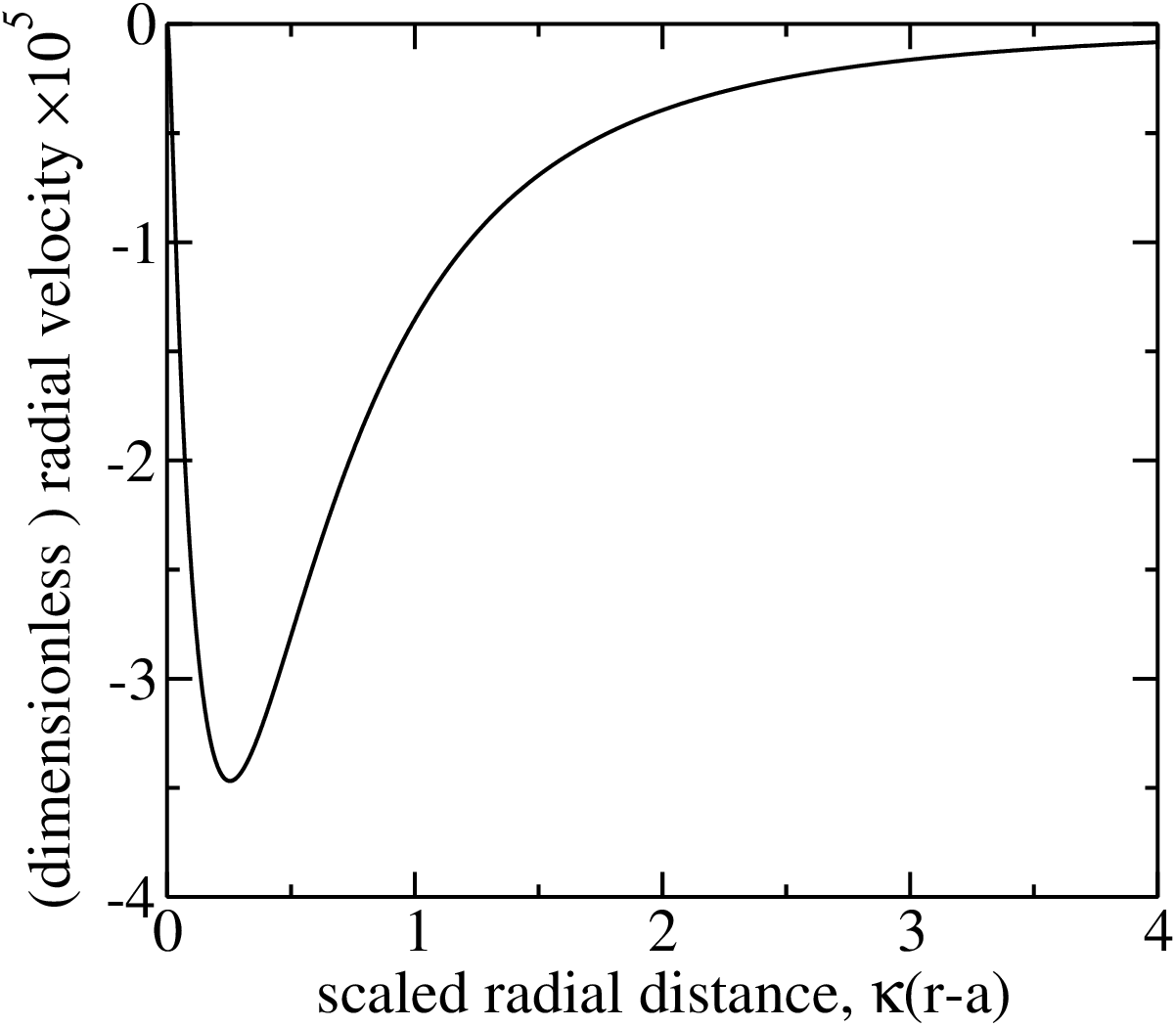} \hspace{1.0cm} \includegraphics[width=6cm]{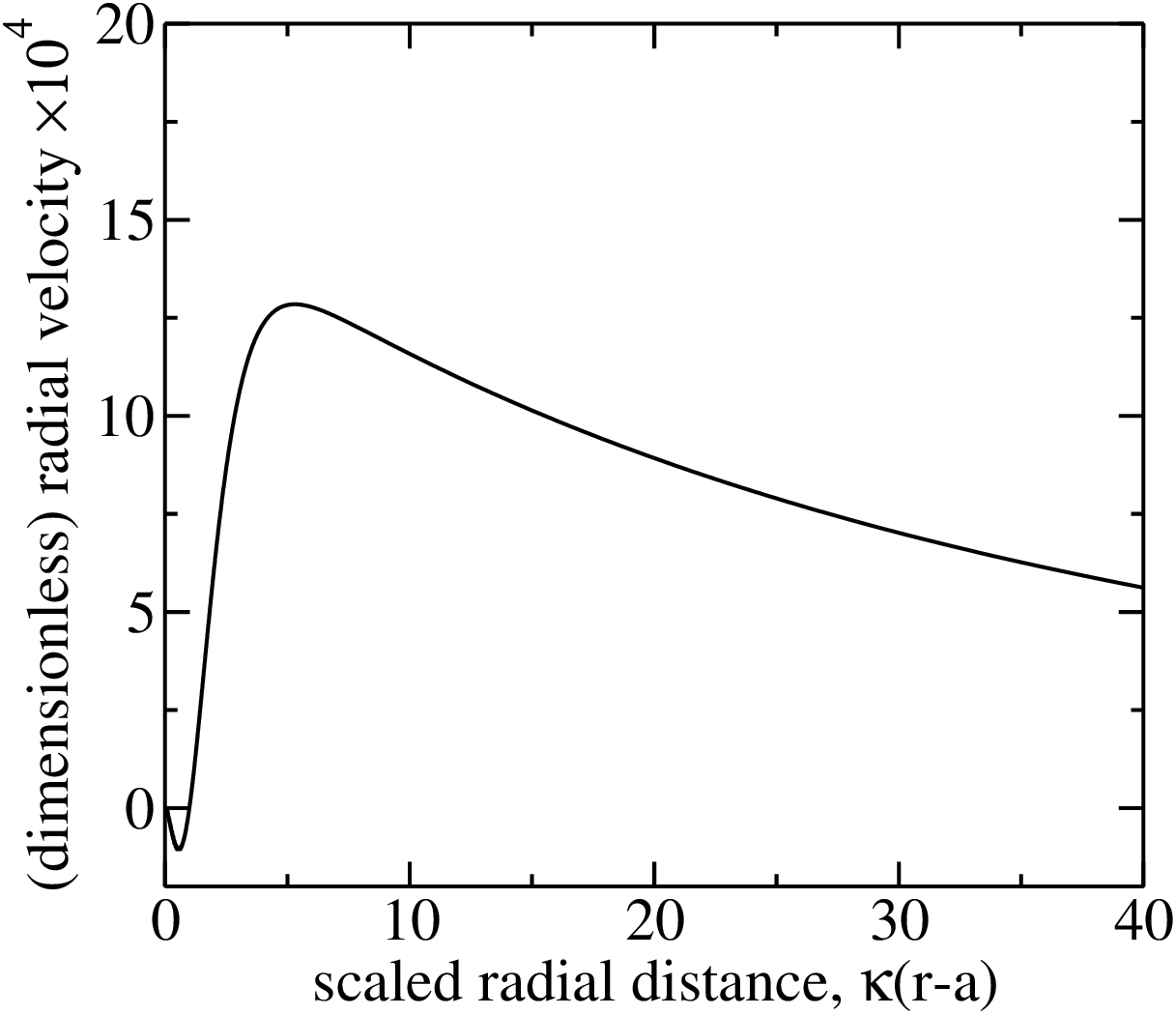}
    \vspace{0.5cm}
  \end{center}
  \caption{\label{fig:perturbedb} The structure of the perturbed
    double layer---under the influence of a bulk electrolyte
    concentration gradient---as revealed by the (dimensionless) radial
    velocity $-2 (h_r / r) 2 I / (u^* a)$ ($u^* = \epsilon_s
    \epsilon_o (\kb T / e)^2 / (\eta a) \approx 5.15 \times
    10^{-3}$~m~s$^{-1}$) (see Eqn.~(\ref{eqn:velocity})) as a function
    of the (scaled) radial position $\kappa (r - a)$. Parameters are
    the same as in figure~\ref{fig:perturbeda}.}
\end{figure}

Asymptotic coefficients are provided in
table~\ref{tab:bfieldalonecoeffs} for a representative polymer gel
with Brinkman screening length $\bsl \approx 0.951$~nm and inclusion
radius $a = 100$~nm; the $\zeta$-potentials and (three) ionic
strengths span experimentally accessible ranges. Here, the sign of the
electrostatic and concentration dipole moments ($D^{B_{k'}}$ and
$C_j^{B_{k'}}$, respectively) and of the far-field flow (as indicated
by $C^{B_{k'}}$) are not straightforward to interpret without
examining the detailed structure of the perturbed double
layer. Representative (dimensionless) functions $\hat{n}_j$ and
$\hat{\psi}(r)$ are shown in figure~\ref{fig:perturbeda} (see
Eqns.~(\ref{eqn:farfield1}) and~(\ref{eqn:farfield2})). Recall, these
characterize the radial variation of the perturbed ion densities and
electrostatic potential. Figure~{\ref{fig:perturbedb}} shows the
corresponding radial fluid velocity, as represented by $-2 h_r / r$
(see Eqn.~(\ref{eqn:velocity})).

From table~\ref{tab:bfieldalonecoeffs}, backward
concentration-gradient-induced flow ($U < 0$ or $C^{B_{k'}} > 0$)
prevails at low to moderate values of $\ka$, with forward flow ($U >
0$) evident at higher values of $\ka$ when $|\zeta|$ is low. Recall,
in the absence of surface charge ($\zeta = 0$), the concentration
dipole moments $C_j^{B_{k'}} = (1/2) a^3$ reflect the (concentration)
polarization necessary for electrolyte to diffuse past impenetrable
inclusions. With surface charge, however, counter-ions (co-ions) at
the forward facing surfaces migrate radially inward (outward) under
the influence of the equilibrium electrostatic potential. The
resulting inner and outer layers of perturbed charge density are
evident in figure~\ref{fig:perturbeda}; these clearly have positive
and negative dipole moments, respectively. Symmetry and
electroneutrality considerations indicate that the sign of the net
electrostatic dipole moment (\ie, the far-field decay of the
electrostatic potential) reflects the moment of the outer layer of
perturbed charge. Accordingly, the electrostatic dipole moments
$D^{B_{k'}}$ in table~\ref{tab:bfieldalonecoeffs} are negative for all
values of $\ka$ and $|\zeta|$.

It is tempting to attribute the direction of the far-field flow to the
sign of the electrostatic dipole moment, whose accompanying electric
field is expected to drive forward flow when acting on the equilibrium
charge density. However, as indicated in
table~\ref{tab:bfieldalonecoeffs}, the far-field flow is most often
backward at low to moderate values of $\ka$. Furthermore, the
direction of the far-field flow is independent of the sign of the net
dipole moment.

In situations where the far-field flow is backward, the inner
perturbed electric field acts on the equilibrium charge density, and
viscous stresses from the prevailing electroosmotic flow propagate to
drive a backward outer flow. In this case, all streamlines are open,
as indicated by the (negative) radial velocity profile in the left
panel of figure~\ref{fig:perturbedb}. However, as seen in the right
panel, situations with a forward far-field flow also have a thin layer
of (relatively slow) backward flow at the inclusion surface. In this
case, closed streamlines prevail within the (thin) equilibrium double
layer. Evidently, electrical forces due to the outer perturbed
electric field acting on the equilibrium charge density dominate
viscous stresses arising from the innermost layer of backward flow.

While the qualitative features of these flows bear similarities to
those underlying the diffusiophoretic motion of charged colloidal
spheres~\citep{Prieve:1987}, the differences are, perhaps, sufficient
to discourage direct comparisons. For example, as seen below, the
direction of the far-field flow in the composites addressed here also
depends on the Brinkman permeability of the intervening
polymer. Furthermore, the diffusiophoretic mobility is also influenced
by an electric field (to maintain zero current density) whose sign
depends on electrolyte asymmetry.

\begin{sidewaystable}
  \begin{center}
    \caption{\label{tab:bfieldalonecoeffs} Asymptotic coefficients
      (see Eqn.~(\ref{eqn:bfieldflux})) for the bulk {\em diffusion}
      of NaCl in a Brinkman medium with charged spherical inclusions:
      $a=100$~nm; $\bsl \approx 0.951$~nm; $T = 25$\degc, $D_1 \approx
      1.33 \times 10^{-9}$m$^2$s$^{-1}$~(Na$^+$); $D_2 \approx 2.03
      \times 10^{-9}$m$^2$s$^{-1}$~(Cl$^-$); $u^* = \epsilon_s
      \epsilon_o (\kb T / e)^2 / (\eta a) \approx 5.15 \times
      10^{-3}$~m~s$^{-1}$.}
    
    \begin{tabular*}{\columnwidth}{@{\extracolsep{\fill}}llllll} \hline
      
       \multicolumn{1}{c}{$\zeta e /(\kb T)$} &
       \multicolumn{1}{c}{$D^{B_{k'}} 2 I e / (\kb T a^3)$} &
       \multicolumn{1}{c}{$C_j^{B_{k'}} / a^3$} &
       \multicolumn{1}{c}{$C^{B_{k'}} 2 I / (u^* a^4)$} &
       \multicolumn{1}{c}{$U / (B_{k'} \phi)$} & \multicolumn{1}{c}{$3
       D^{B_{k'}}/a^3$}\\ & & \multicolumn{1}{c}{($j=1,2$)} & &
       \multicolumn{1}{c}{$=-3C^{B_{k'}}/a^3$} & \\ & & & &
       \multicolumn{1}{c}{((nm~s$^{-1}$)/(\M~cm$^{-1}$))} &
       \multicolumn{1}{c}{(V/(\M))}\\ \hline

$\ka=  1$ & $I=9.25\times 10^{-6}$~\M \\ \hline

$-1$ & $-4.16\times 10^{0}$ & $+3.82\times 10^{-1}$ & $1.51\times 10^{-5}$ & $-1.26\times 10^{2}$ & $-1.73\times 10^{4}$ \\ 
$-2$ & $-8.16\times 10^{0}$ & $+5.71\times 10^{-2}$ & $5.21\times 10^{-5}$ & $-4.35\times 10^{2}$ & $-3.40\times 10^{4}$ \\ 
$-4$ & $-1.49\times 10^{1}$ & $-8.66\times 10^{-1}$ & $1.14\times 10^{-4}$ & $-9.53\times 10^{2}$ & $-6.20\times 10^{4}$ \\ 
$-6$ & $-1.91\times 10^{1}$ & $-1.60\times 10^{0}$  & $1.14\times 10^{-4}$ & $-9.52\times 10^{2}$ & $-7.95\times 10^{4}$ \\ 
$-8$ & $-2.11\times 10^{1}$ & $-1.96\times 10^{0}$  & $9.77\times 10^{-5}$ & $-8.16\times 10^{2}$ & $-8.79\times 10^{4}$ \\ \hline

$\ka= 10$ & $I=9.25\times 10^{-4}$~\M \\ \hline

$-1$ & $-2.92\times 10^{-1}$ & $+4.74\times 10^{-1}$ & $2.54\times 10^{-5}$ & $-2.12\times 10^{0}$ & $-1.22\times 10^{1}$ \\ 
$-2$ & $-6.04\times 10^{-1}$ & $+3.98\times 10^{-1}$ & $9.99\times 10^{-5}$ & $-8..4\times 10^{0}$ & $-2.52\times 10^{1}$ \\ 
$-4$ & $-1.29\times 10^{0}$  & $+1.42\times 10^{-1}$ & $3.06\times 10^{-4}$ & $-2.56\times 10^{1}$ & $-5.37\times 10^{1}$ \\ 
$-6$ & $-1.88\times 10^{0}$  & $-1.20\times 10^{-1}$ & $4.12\times 10^{-4}$ & $-3.44\times 10^{1}$ & $-7.84\times 10^{1}$ \\ 
$-8$ & $-2.23\times 10^{0}$  & $-2.81\times 10^{-1}$ & $4.26\times 10^{-4}$ & $-3.56\times 10^{1}$ & $-9.29\times 10^{1}$ \\ \hline

$\ka=100$ & $I=9.25\times 10^{-2}$~\M \\ \hline

$-1$ & $-3.08\times 10^{-2}$ & $4.96\times 10^{-1}$ & $-3.01\times 10^{-4}$ & $+2.52\times 10^{-1}$ & $-1.28\times 10^{-2}$ \\ 
$-2$ & $-6.85\times 10^{-2}$ & $4.85\times 10^{-1}$ & $-7.71\times 10^{-4}$ & $+6.44\times 10^{-1}$ & $-2.85\times 10^{-2}$ \\ 
$-4$ & $-1.97\times 10^{-1}$ & $4.28\times 10^{-1}$ & $-6.87\times 10^{-4}$ & $+5.74\times 10^{-1}$ & $-8.19\times 10^{-2}$ \\ 
$-6$ & $-4.49\times 10^{-1}$ & $3.04\times 10^{-1}$ & $-3.05\times 10^{-5}$ & $+2.55\times 10^{-2}$ & $-1.87\times 10^{-1}$ \\ 
$-8$ & $-8.24\times 10^{-1}$ & $1.18\times 10^{-1}$ & $+3.15\times 10^{-4}$ & $-2.63\times 10^{-1}$ & $-3.43\times 10^{-1}$ \\ \hline
      
    \end{tabular*}
  \end{center}
\end{sidewaystable}

\begin{figure}
  \begin{center}
    \vspace{1.0cm} \includegraphics[width=6cm]{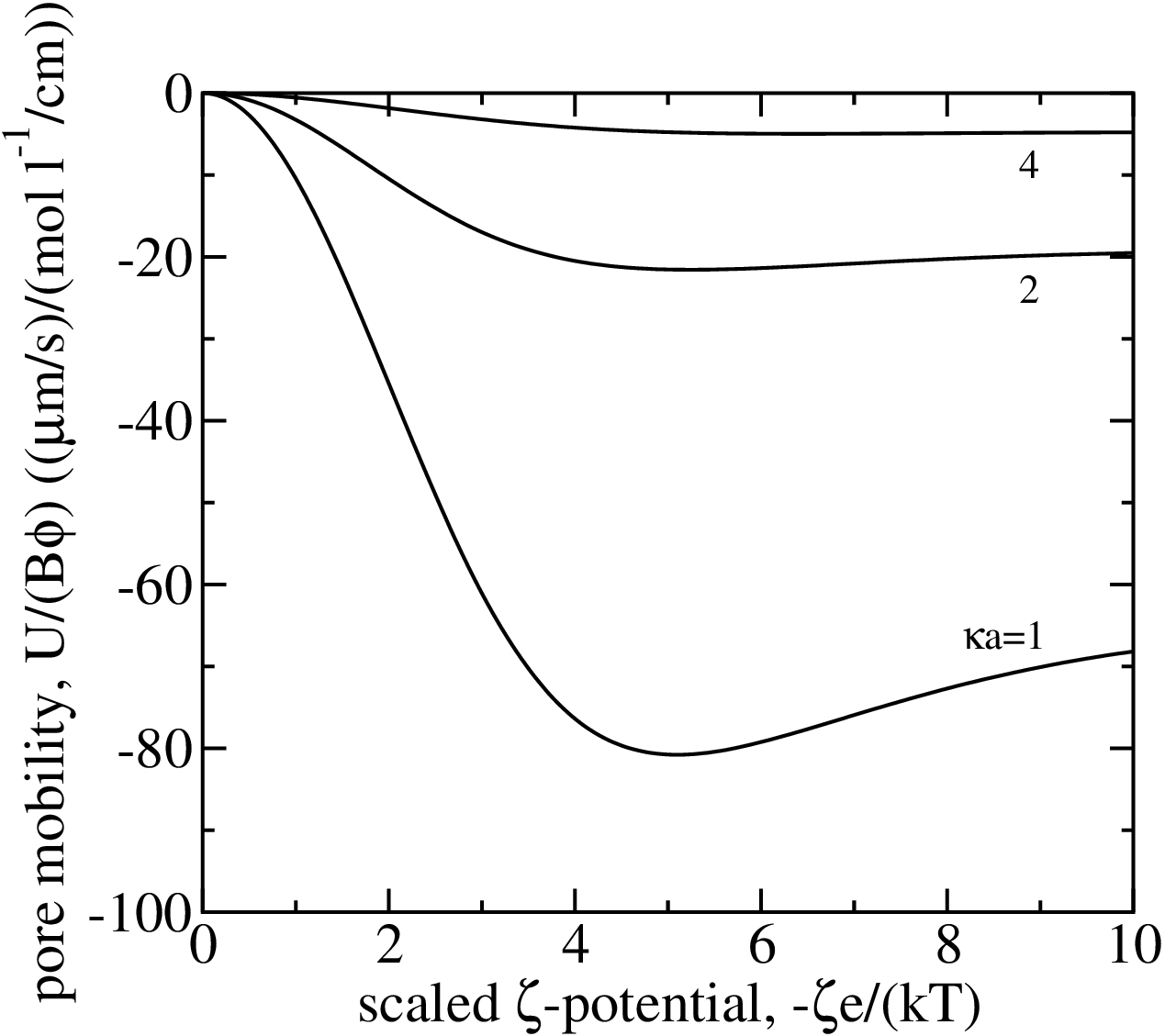}
    \vspace{0.5cm}
  \end{center}
  \caption{\label{fig:mobilityBBSL=9.98NM} The incremental pore
    mobility $U / (B_{k'} \phi) = -3 C^{B_{k'}} / a^3$ as a function
    of the (scaled) $\zeta$-potential $\zeta e/(\kb T)$ for (scaled)
    reciprocal double-layer thicknesses $\ka = 1$, 2 and 4: aqueous
    NaCl at $T = 25$\degc; $a = 100$~nm; $\bsl \approx 9.95$~nm. An
    average concentration gradient is applied in the absence of
    average pressure and electrostatic potential gradients.}
\end{figure}

\begin{figure}
  \begin{center}
    \vspace{1.0cm} \includegraphics[width=6cm]{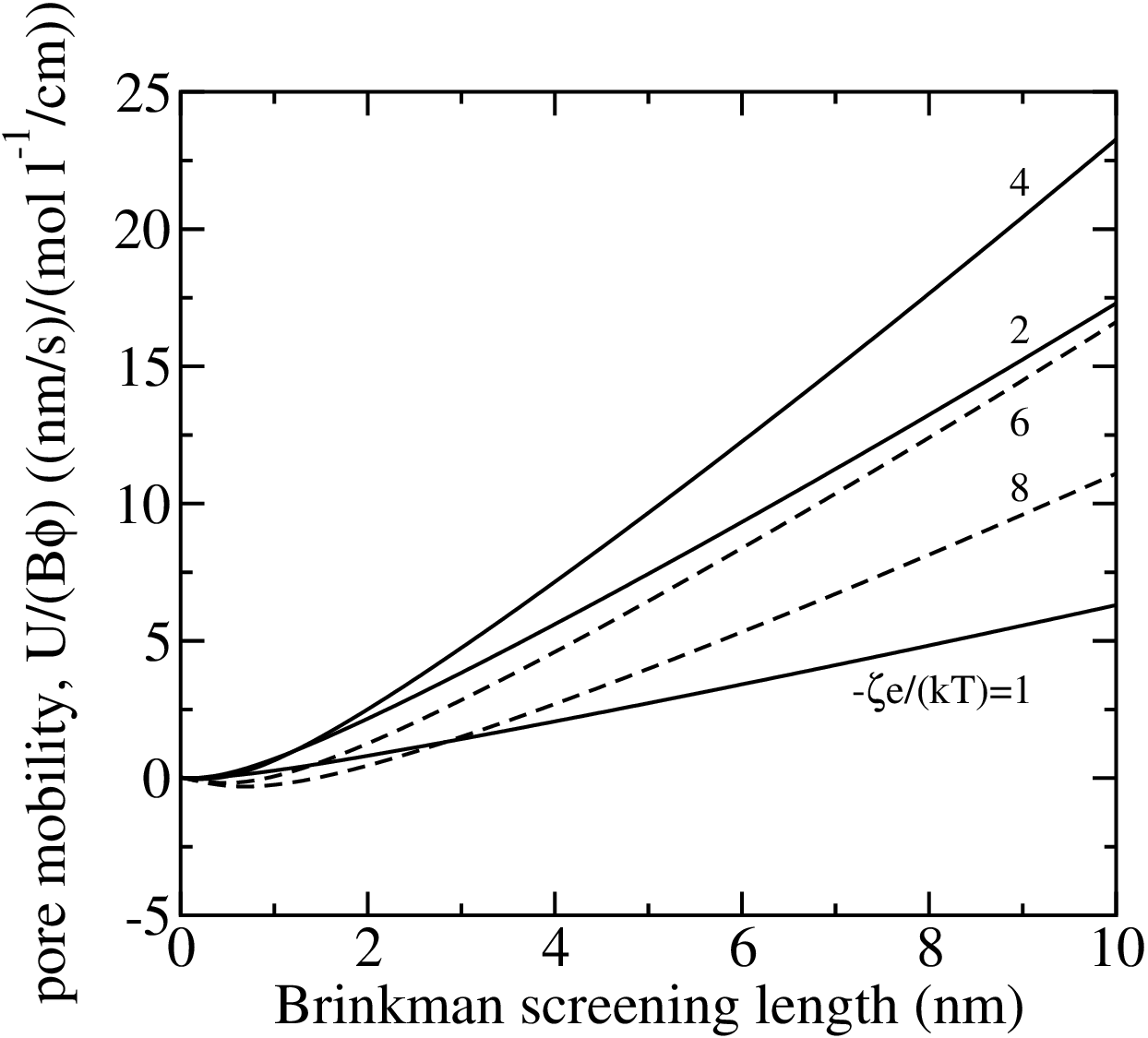}
    \vspace{0.5cm}
  \end{center}
  \caption{\label{fig:mobilityBka=100} The incremental pore mobility
    $U / (B_{k'} \phi) = - 3 C^{B_{k'}} / a^3$ as a function of the
    Brinkman screening length $\bsl$ for various (scaled)
    $\zeta$-potentials $-\zeta e/(\kb T) = 1$, $2$, $4$ ({\em solid
    lines}) $6$ and $8$ ({\em dashed lines}): aqueous NaCl at $T =
    25$\degc; $a = 100$~nm; $\ka = 100$; $I \approx 0.0925$~\M. An
    average concentration gradient is applied in the absence of
    average pressure and electrostatic potential gradients.}
\end{figure}

For a composite with a relatively large Brinkman screening length
$\bsl \approx 9.95$~nm, the concentration-gradient-induced pore
velocities shown in figure~\ref{fig:mobilityBBSL=9.98NM} are low. As
expected, the incremental pore mobility increases with the surface
charge (or $\zeta$-potential) {\em and} Darcy permeability. The
mobility is shown in figure~\ref{fig:mobilityBka=100} as a function of
the Brinkman screening length for various $\zeta$-potentials, with an
ionic strength $I \approx 0.0925$~\M \ yielding $\ka = 100$. The
direction of the concentration-gradient-induced flow is consistent
with the discussion of table~\ref{tab:bfieldalonecoeffs}. Increasing
the gel permeability strengthens the (forward flowing) electroosmotic
flow in the outer region of the perturbed double layers and, hence,
increases the bulk convective flow.

Contributions to the incremental fluxes are provided in
table~\ref{tab:bfieldaloneincs} for the composite whose asymptotic
coefficients are listed in table~\ref{tab:bfieldalonecoeffs}. The
first term on the right-hand side of Eqn.~(\ref{eqn:bfieldincrements})
(also columns 2 and 3 of table~\ref{tab:bfieldaloneincs}) is due to
electromigration arising from concentration-gradient-induced
electrical polarization. Polarization of (negatively) charged
inclusions evidently enhances the effective diffusivity of the
counter-ion (Na$^+$) and attenuates that of the co-ion (Cl$^-$). The
second term in Eqn.~(\ref{eqn:bfieldincrements}) (column 4 of the
table) is due to diffusion, and arises from perturbations to the ion
concentration gradients. As expected, this attenuates the diffusive
flux when the $\zeta$-potential is moderate, and approaches the
Maxwell value $\Delta_{j,d}^{B_{k'}} = -3/2$ as $|\zeta| \rightarrow
0$. The last term in Eqn.~(\ref{eqn:bfieldincrements}) (also column 5)
is due to electroosmotic flow (convection). This diminishes the fluxes
of both ions, but its contribution is relatively small. Summing all
three contributions (last two columns) shows that the average
diffusive flux of the co-ion (Cl$^-$) is attenuated by the inclusions,
most significantly at low ionic strength when electrostatic screening
of the surface charge is weak. Similarly, the flux of the counter-ion
(Na$^+$) is enhanced because of electromigration. Again, when
electrical forces are weak, the inclusions simply hinder diffusion and
$\Delta_j^{B_{k'}} \rightarrow -3/2$ as $|\zeta| \rightarrow 0$.

\begin{sidewaystable}
  \small 
  \begin{center}
    \caption{\label{tab:bfieldaloneincs} Scaled asymptotic
      coefficients and microscale incremental contributions (see text)
      to average diffusion of NaCl in a Brinkman medium with charged
      spherical inclusions: $a=100$~nm; $\bsl \approx 0.951$~nm; $T =
      25$\degc; $D_1 \approx 1.33 \times 10^{-9}$m$^2$s$^{-1}$~(Na$^+$);
      $D_2 \approx 2.03 \times 10^{-9}$m$^2$s$^{-1}$~(Cl$^-$).}
    \begin{tabular*}{\columnwidth}{@{\extracolsep{\fill}}lllllllllll} \hline
      \multicolumn{1}{c}{ $\zeta e /(\kb T)$} &
      \multicolumn{1}{c}{$\Delta_{1,e}^{B_{k'}}$~(Na$^+$) ($=-\Delta_{2,e}^{B_{k'}}$)} &
      \multicolumn{1}{c}{$\Delta_{j,d}^{B_{k'}}$ ($j=1,2$)}&
      \multicolumn{1}{c}{$\Delta_{1,c}^{B_{k'}}$~(Na$^+$)} &
      \multicolumn{1}{c}{$\Delta_{2,c}^{B_{k'}}$~(Cl$^-$)} & 
      \multicolumn{1}{c}{$\Delta_1^{B_{k'}}$~(Na$^+$)} & 
      \multicolumn{1}{c}{$\Delta_2^{B_{k'}}$~(Cl$^-$)}\\ \hline
      
      $\ka=  1$ & $I=9.25\times 10^{-6}$~\M \\ \hline
      
      $-1$ & $6.25\times 10^{0}$ & $-1.15\times 10^{0}$  & $-8.76\times 10^{-6}$ & $-5.74\times 10^{-6}$ & $5.10\times 10^{0}$ & $-7.39\times 10^{0}$\\
      $-2$ & $1.22\times 10^{1}$ & $-1.71\times 10^{-1}$ & $-3.02\times 10^{-5}$ & $-1.98\times 10^{-5}$ & $1.21\times 10^{1}$ & $-1.24\times 10^{1}$\\
      $-4$ & $2.23\times 10^{1}$ & $+2.60\times 10^{0}$  & $-6.61\times 10^{-5}$ & $-4.34\times 10^{-5}$ & $2.49\times 10^{1}$ & $-1.97\times 10^{1}$\\
      $-6$ & $2.86\times 10^{1}$ & $+4.79\times 10^{0}$  & $-6.61\times 10^{-5}$ & $-4.33\times 10^{-5}$ & $3.34\times 10^{1}$ & $-2.39\times 10^{1}$\\
      $-8$ & $3.17\times 10^{1}$ & $+5.89\times 10^{0}$  & $-5.66\times 10^{-5}$ & $-3.71\times 10^{-5}$ & $3.75\times 10^{1}$ & $-2.58\times 10^{1}$\\ \hline
      
      $\ka= 10$ & $I=9.25\times 10^{-4}$~\M \\ \hline
      
      $-1$ & $4.38\times 10^{-1}$ & $-1.42\times 10^{0}$  & $-1.47\times 10^{-5}$ & $-9.65\times 10^{-6}$ & $-9.84\times 10^{-1}$ & $-1.86\times 10^{0}$\\
      $-2$ & $9.07\times 10^{-1}$ & $-1.19\times 10^{0}$  & $-5.79\times 10^{-5}$ & $-3.80\times 10^{-5}$ & $-2.88\times 10^{-1}$ & $-2.10\times 10^{0}$\\
      $-4$ & $1.94\times 10^{0}$  & $-4.25\times 10^{-1}$ & $-1.77\times 10^{-4}$ & $-1.16\times 10^{-4}$ & $+1.51\times 10^{0}$  & $-2.36\times 10^{0}$\\
      $-6$ & $2.82\times 10^{0}$  & $+3.61\times 10^{-1}$ & $-2.39\times 10^{-4}$ & $-1.57\times 10^{-4}$ & $+3.19\times 10^{0}$  & $-2.46\times 10^{0}$\\
      $-8$ & $3.35\times 10^{0}$  & $+8.42\times 10^{-1}$ & $-2.47\times 10^{-4}$ & $-1.62\times 10^{-4}$ & $+4.19\times 10^{0}$  & $-2.50\times 10^{0}$\\ \hline
      
      $\ka=100$ & $I=9.25\times 10^{-2}$~\M \\ \hline
      
      $-1$ & $4.63\times 10^{-2}$ & $-1.49\times 10^{0}$  & $+1.75\times 10^{-4}$ & $+1.14\times 10^{-4}$ & $-1.44\times 10^{0}$  & $-1.54\times 10^{0}$\\
      $-2$ & $1.03\times 10^{-1}$ & $-1.45\times 10^{0}$  & $+4.47\times 10^{-4}$ & $+2.93\times 10^{-4}$ & $-1.35\times 10^{0}$  & $-1.56\times 10^{0}$\\
      $-4$ & $2.95\times 10^{-1}$ & $-1.28\times 10^{0}$  & $+3.98\times 10^{-4}$ & $+2.61\times 10^{-4}$ & $-9.88\times 10^{-1}$ & $-1.58\times 10^{0}$\\
      $-6$ & $6.74\times 10^{-1}$ & $-9.13\times 10^{-1}$ & $+1.77\times 10^{-5}$ & $+1.16\times 10^{-5}$ & $-2.39\times 10^{-1}$ & $-1.59\times 10^{0}$\\
      $-8$ & $1.24\times 10^{0}$  & $-3.53\times 10^{-1}$ & $-1.82\times 10^{-4}$ & $-1.20\times 10^{-4}$ & $+8.83\times 10^{-1}$ & $-1.59\times 10^{0}$\\ \hline
      
    \end{tabular*}
  \end{center}
\end{sidewaystable}

\begin{figure}
  \begin{center}
    \vspace{1.0cm} \includegraphics[width=6cm]{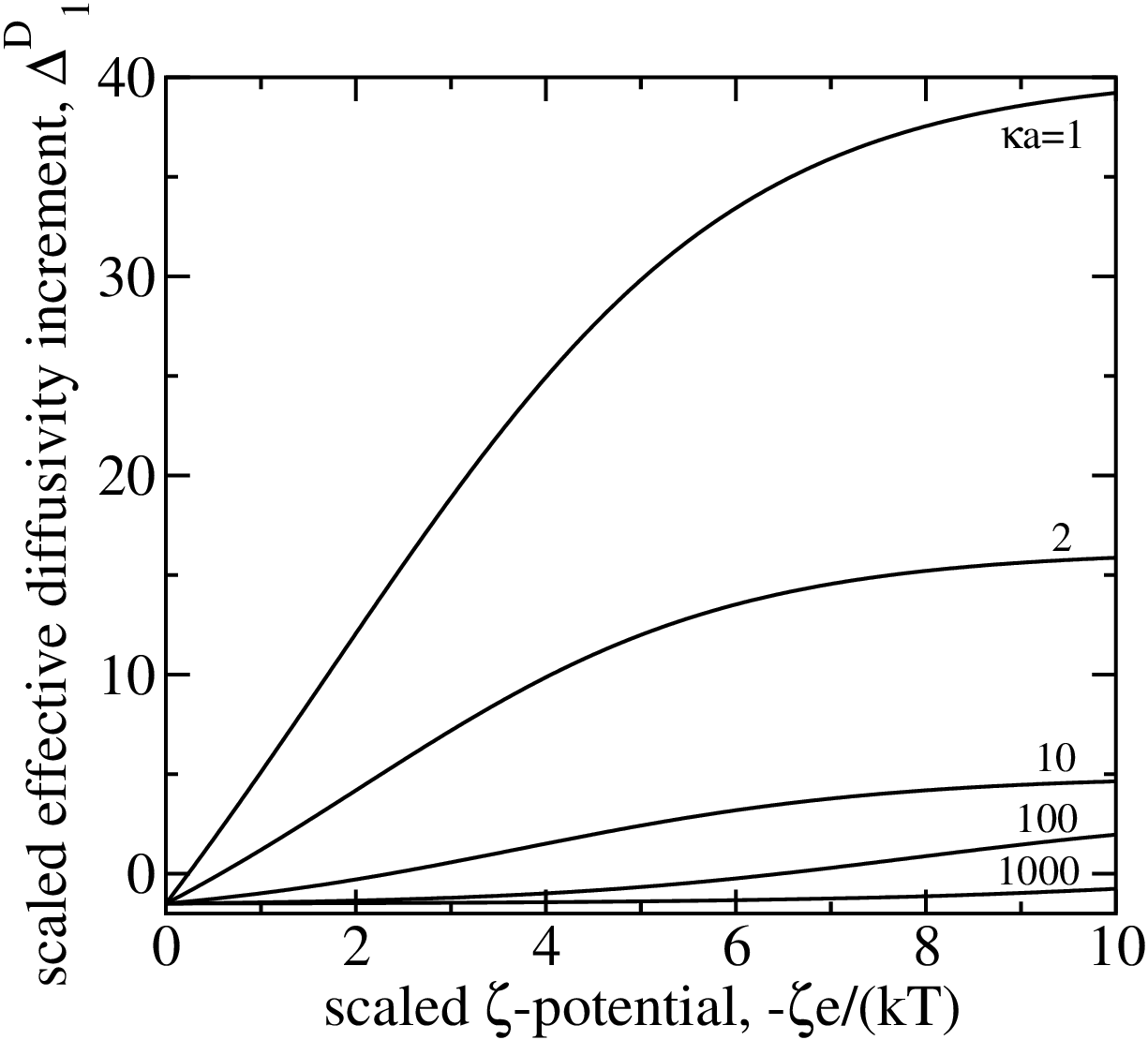}
    \hspace{1cm} \includegraphics[width=6cm]{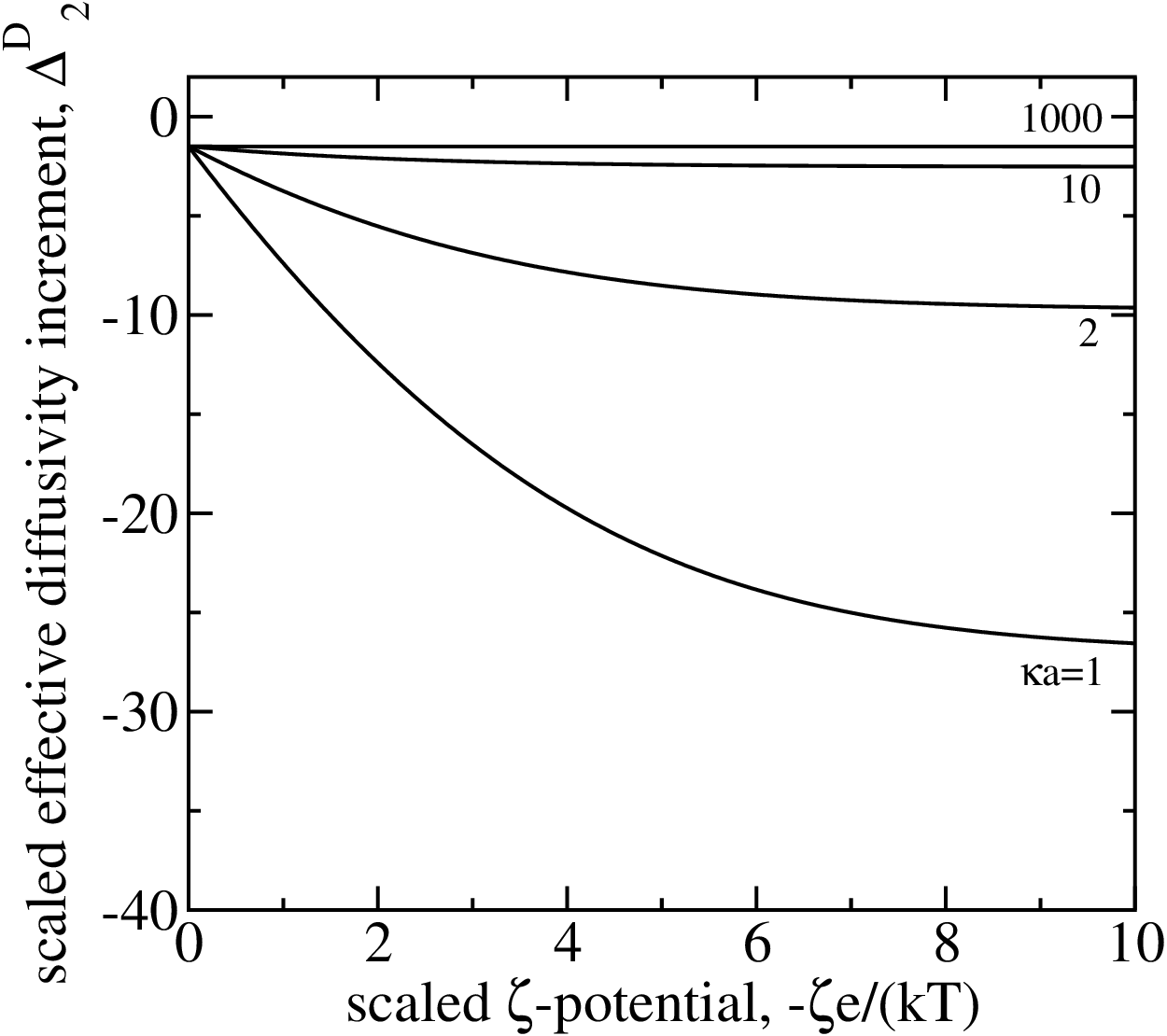}
    \vspace{0.5cm}
  \end{center}
  \caption{\label{fig:diffincB} The (scaled) effective diffusivity
    increments $\Delta_j^D = (D_j^* / D_j - 1)/ \phi$ for Na$^+$ ({\em
    left panel}) and Cl$^-$ ({\em right panel}) as a function of the
    (scaled) $\zeta$-potential $\zeta e / (\kb T)$ with various
    (scaled) reciprocal double-layer thicknesses $\ka = 1$, 2, 10, 100
    and 1000 (aqueous NaCl at $T = 25$\degc \ with $a = 100$~nm). Note
    that results are insensitive to the permeability $\bsl^2$ of the
    polymer gel.}
\end{figure}

Effective diffusivity increments are shown in
figure~\ref{fig:diffincB} as a function of the $\zeta$-potential for
various values of $\ka$. Clearly, increasing the surface charge
increases the effective diffusivity of the counter-ion (left panel)
and decreases the diffusivity of the co-ion (right panel). Note that
the calculations are insensitive to the gel permeability, so these
results are applicable to a variety of composites with negatively
charged inclusions and NaCl electrolyte. In general, all bulk
properties that reflect concentration and electrical polarization are
independent of the Darcy permeability.

\subsection{Bulk diffusion and electromigration} \label{sec:diffusionandelectromigration}

Having examined the application of an electrolyte concentration
gradient in the absence of an electric field, let us now consider the
electric field strength and concentration gradient that together yield
zero current density. This ensures that transport of ions from one
reservoir to another (with different electrolyte concentration)
maintains electrical neutrality across a membrane. As expected from
the previous section, an electric field is necessary to compensate for
the lower (higher) diffusive flux of the less (more) mobile ion.

The analysis below is limited to situations where the bulk electrolyte
concentration gradient is weak, permitting macroscale variations in
the bulk equilibrium electrolyte concentration to be
neglected. Complications that arise from macroscale variations in ion
density, which are necessary to satisfy (bulk) continuity and
electroneutrality constraints, will be addressed in future work; these
are discussed briefly in~\S\ref{sec:summary}.

For the simple case involving the concentration gradient of a single
$z$-$z$ electrolyte ($B_j = B_{k'}$, $N=2$ and $M=1$), setting the
current density (Eqn.~(\ref{eqn:diffusioncurrent})) to zero gives
\begin{equation} \label{eqn:gradientpotential}
  \vect{E} K^\infty = \vect{B}_{k'} \sum_{j=1}^{N=2} z_j e D_j (1 +
  \phi \Delta^E) + O(\phi^2),
\end{equation}
where
\begin{equation} \label{eqn:fieldinc}
  \Delta^E \approx -(3 / a^3) [ D^{B_{k'}}
    \frac{K^\infty}{\sum_{j'=1}^{N=2} z_{j'} e D_{j'}} +
    C_{j}^{B_{k'}} + D^E + C_{j}^E \frac{\sum_{j'=1}^{N=2} z_{j'} e
      D_{j'}}{K^\infty}] + O(\phi)
\end{equation}
is termed the (dimensionless) {\em electric-field increment}. Note
that $C_j^E$ and $C_j^{B_{k'}}$ are independent of $j$, so $\Delta^E$
is independent of the sum in Eqn.~(\ref{eqn:gradientpotential}).

Similarly to the concentration-gradient-induced current density
(Eqn.~\ref{eqn:diffusioncurrent}), the $O(1)$ average electric field
is zero for $z$-$z$ electrolytes whose ions have equal mobilities. It
is important to note that the $O(\phi)$ contribution is not zero,
however, even for perfectly symmetrical electrolytes. This follows
from the influence of charged inclusions on the effective ion
diffusivities, and emerges from Eqn.~(\ref{eqn:fieldinc}) through the
term involving the gradient-induced electrostatic dipole strength
$D^{B_{k'}}$. Therefore, for a $z$-$z$ electrolyte whose ions have
equal diffusivities $D_j$, Eqn.~(\ref{eqn:gradientpotential})
simplifies to
\begin{equation} \label{eqn:simpler}
  \vect{E} = - \phi (3 / a^3) D^{B_{k'}} \vect{B}_{k'} + O(\phi^2),
\end{equation}
showing that the average electric field reflects a sum of
concentration-gradient-induced electrostatic dipole moments. In an
homogeneous membrane, the membrane diffusion potential $\Delta \psi =
\psi(z=L) - \psi(0) = - E L = \phi (3 / a^3) D^{B_{k'}} B_{k'} L$ is
simply proportional to the concentration differential $B_{k'} L$.

\begin{figure}
  \begin{center}
    \vspace{1.0cm} \includegraphics[width=6cm]{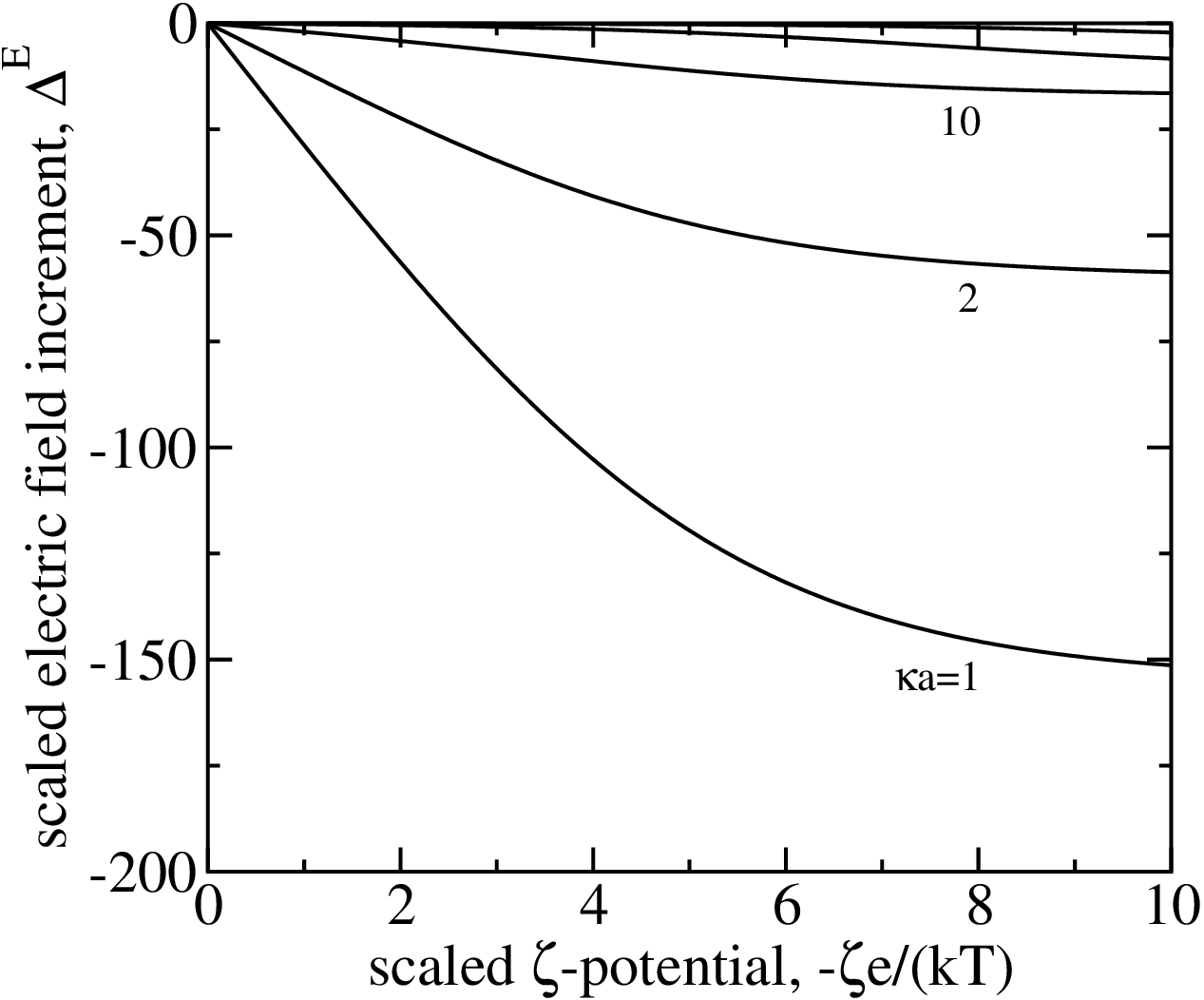}
    \hspace{1cm} \includegraphics[width=6cm]{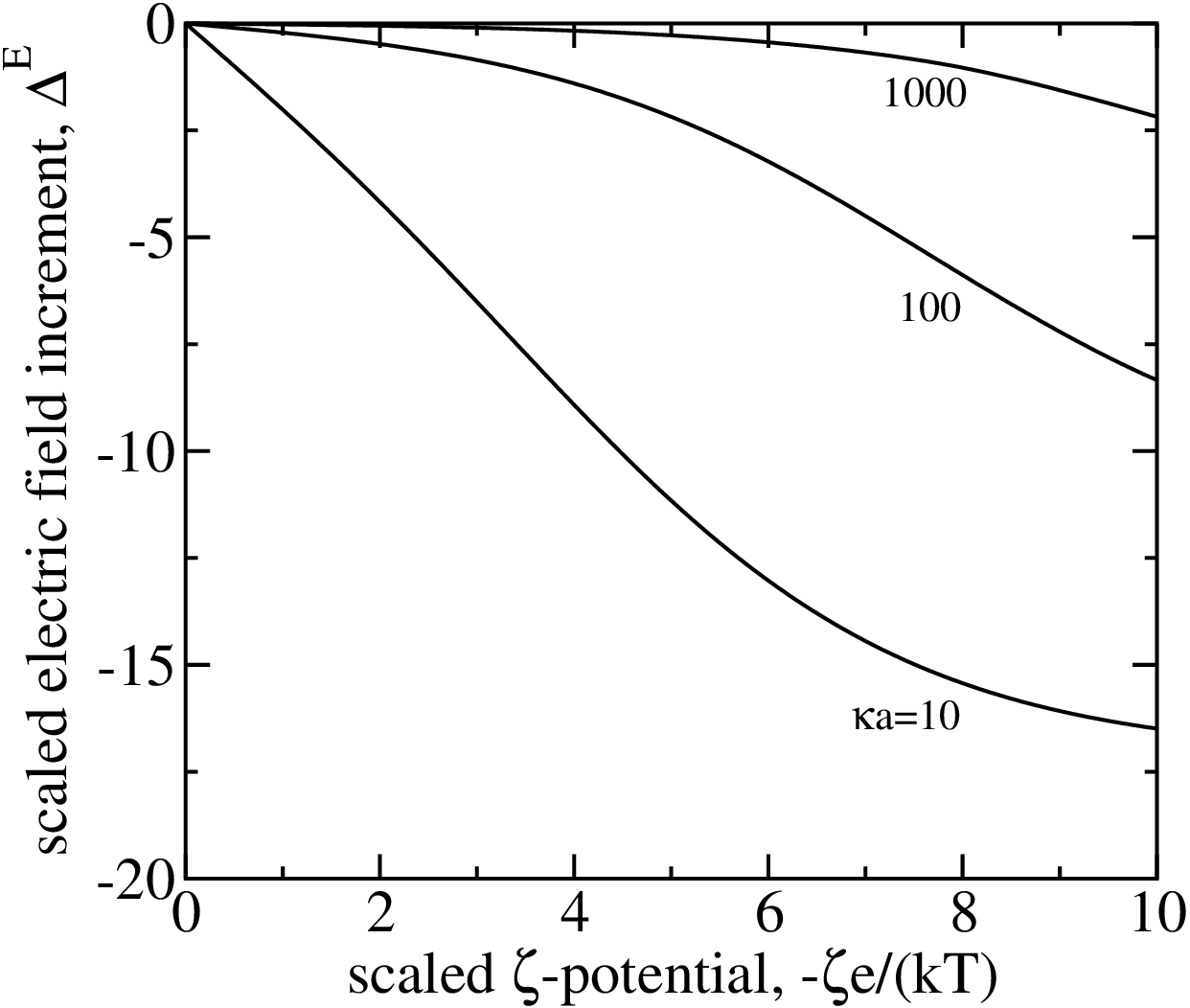}
    \vspace{0.5cm}
  \end{center}
  \caption{\label{fig:membpot} The (scaled) electric-field increment
    $\Delta^E$ (see
    Eqns.~(\ref{eqn:gradientpotential})--(\ref{eqn:fieldinc})) as a
    function of the (scaled) $\zeta$-potential $-\zeta e/(\kb T)$ for
    various (scaled) reciprocal double-layer thicknesses $\ka = 1$, 2,
    10, 100 and 1000: aqueous NaCl at $T = 25$\degc; $a =
    100$~nm. Note that the results are insensitive to the permeability
    $\bsl^2$ of the polymer gel.}
\end{figure}

The electric-field increment is shown in figure~\ref{fig:membpot} as a
function of the $\zeta$-potential for various bulk ionic strengths
yielding $\ka$ in the range 1--$10^3$. Again, because the increment
reflects electrical and concentration polarization, the calculations
are insensitive to flow and, hence, Darcy permeability. Therefore, the
results are applicable to a variety of composites with negatively
charged inclusions and NaCl electrolyte. Note that the increment is
large at low ionic strength (small $\ka$) when the $\zeta$-potential
is high, so the particle contribution may be comparable to or greater
than that of the electrolyte alone. Further, when $\Delta^E < 0$,
there exists an inclusion volume fraction
\begin{equation}
  \phi^* = - 1 / \Delta^E
\end{equation}
that yields zero electric field strength with any (weak) bulk
electrolyte concentration gradient. Clearly, $\phi^*$ must be less
than one to be consistent with assumptions underlying the theory. In
turn, this limits the accuracy of such a prediction to situations
where $-\Delta^E \sim \phi^{-1}$ ($\phi \ll 1$), which is achieved at
low ionic strength (\eg, $\ka = 1$ in figure~\ref{fig:membpot}).

\section{Summary} \label{sec:summary}

This paper demonstrates that a weak electrolyte concentration gradient
applied across a permeable membrane hosting immobile charged
inclusions induces bulk electroosmotic (or ``diffusioosmotic'')
flow. The direction of this flow may be up or down the concentration
gradient, depending on the surface charge, bulk ionic strength, and
permeability of the polymer gel. The inclusions also influence the ion
fluxes or, equivalently, the effective ion diffusion coefficients. In
short, the net counter-ion flux is enhanced by electromigration within
the diffuse double layers.

Steady or quasi-steady transport across a membrane often takes place
with zero current density, so, in practice, an external electric
field---or an electric field from a macroscale redistribution of space
charge---is necessary to maintain electrical neutrality. This work
established the proportionality between this electric field and the
imposed concentration gradient. It is tempting to attribute the
strength of the electric field to electrolyte asymmetry alone,
concluding, for example, that a perfectly symmetrical electrolyte will
not produce a membrane diffusion potential. However, it was
demonstrated that charged inclusions break the (effective) symmetry,
so even a perfectly symmetrical electrolyte will yield a membrane
diffusion potential. For NaCl (a moderately asymmetric $z$-$z$
electrolyte) in the presence of negatively charged inclusions, the
electric-field increment is negative, so the diffusion potential of
the composite is lower than in the absence of inclusions.

It should be emphasized that the present theory is limited to
homogeneous membranes. Therefore, with a macroscopic length scale $L$
(\eg, membrane thickness), the differential concentrations $\Delta
n_j^\infty = B_j L$ should be small relative to the bulk
concentrations $n_j^\infty$. Even in the absence of inclusions, the
bulk electrolyte fluxes are not straightforward to calculate without
adopting one of the two following approximations. One consequence of
electrical neutrality is that the electric field (\eg, with
unidirectional transport) must be uniform. Similarly, ion conservation
at steady state demands constant ion fluxes. However, solving the ion
transport equations under these conditions leads to ion concentration
fields that do not yield an electrically neutral bulk. This is the
scenario that underlies the Goldman-Hodgkin-Katz (GHK) equation, which
is well known in membrane biology~\citep{Sten-Knudsen:2002}. An
alternative approach constrains the ion concentration fields to ensure
electrical neutrality. Then, however, the (bulk) ion fluxes are not
uniform and, hence, the ion concentrations cannot be steady. This
scenario underlies the Henderson equation, which is well known in
electrochemistry and electrochemical
engineering~\citep{Sten-Knudsen:2002}. It is interesting to note that
both theories yield the same membrane diffusion potential $\Delta \psi
= \psi(z=L) - \psi(0) = - E L$ when the equations are linearized for
small electrostatic potentials $\psi < \kb T / e$. Given the
difficulties in solving coupled electromigration and diffusion
(electro-diffusion) in the absence of inclusions, the problem with
charged inclusions seems intractable at present. This work will
hopefully stimulate further work in this area.

\begin{acknowledgments}
  Supported by the Natural Sciences and Engineering Research Council
  of Canada (NSERC), through grant number 204542, and the Canada
  Research Chairs program (Tier II). The author thanks S. Omanovic and
  J. Vera (McGill University) and I. Ispolatov (University of
  Santiago) for helpful discussions related to this work.
\end{acknowledgments}

\bibliography{../../../bibliographies/global}

\begin{thebibliography}{10}

\bibitem{Anderson:1989}
J.~L. Anderson.
\newblock Colloidal transport by interfacial forces.
\newblock {\em Ann. Rev. Fluid Mech.}, 21:61--99, 1989.

\bibitem{Booth:1950}
F.~Booth.
\newblock The cataphoresis of spherical, solid non-conducting particles in a
  symmetrical electrolyte.
\newblock {\em Proc. Roy. Soc. Lond.}, 203:533--551, 1950.

\bibitem{Brinkman:1947}
H.~C. Brinkman.
\newblock A calculation of the viscous force exerted by a flowing fluid on a
  dense swarm of particles.
\newblock {\em Appl. Sci. Res. A}, 1:27--34, 1947.

\bibitem{Debye:1948}
P.~Debye and A.~M. Bueche.
\newblock Intrinsic viscosity, diffusion, and sedimentation rate of polymers in
  solution.
\newblock {\em J. Chem. Phys.}, 16(6):573--578, 1948.

\bibitem{Dukhin:1974}
S.~S. Dukhin and B.~V. Derjaguin.
\newblock {\em Electrokinetic Phenomena}, volume~7 of {\em Surface \& Colloid
  Science}.
\newblock Wiley, New York, 1974.

\bibitem{Hagedorn:2005}
R.~Hagedorn, T.~Schnelle, T.~M\"{u}ller, I.~Scholz, K.~Lange, and M.~Reh.
\newblock Electrophoresis in gel channels.
\newblock {\em Electrophoresis}, 26:2495--2502, 2005.

\bibitem{Hill:2004a}
R.~J. Hill.
\newblock Hydrodynamics and electrokinetics of spherical liposomes with
  coatings of terminally anchored poly(ethylene glycol): Numerically exact
  electrokinetics with self-consistent mean-field polymer.
\newblock {\em Phys. Rev. E}, 70:051046, 2004.

\bibitem{Hill:2005b}
R.~J. Hill.
\newblock Transport in polymer-gel composites: {Theoretical} methodology and
  response to an electric field.
\newblock {\em J. Fluid Mech. (In press)}, 2005.

\bibitem{Hill:2005a}
R.~J. Hill and D.~A. Saville.
\newblock `{Exact}' solutions of the full electrokinetic model for soft
  spherical colloids: Electrophoretic mobility.
\newblock {\em Colloids and Surfaces A: Physicochem. Eng. Aspects}, 267:31--49,
  2005.

\bibitem{Hill:2003a}
R.~J. Hill, D.~A. Saville, and W.~B. Russel.
\newblock Electrophoresis of spherical polymer-coated colloidal particles.
\newblock {\em J. Colloid Interface Sci.}, 258:56--74, 2003.

\bibitem{Koch:1999}
D.~L. Koch and A.~S. Sangani.
\newblock Particle pressure and marginal stability limits for a homogeneous
  monodisperse gas fluidized bed: {Kinetic} theory and numerical simulations.
\newblock {\em J. Fluid Mech.}, 400:229--263, 1999.

\bibitem{Lakshminarayanaiah:1969}
N.~Lakshminarayanaiah.
\newblock {\em Transport Phenomena in Membranes}.
\newblock Academic Press, 1969.

\bibitem{Levine:1983}
S.~Levine, K.~Levine, K.~A. Sharp, and D.~E. Brooks.
\newblock Theory of the electrokinetic behavior of human erythrocytes.
\newblock {\em Biophys. J.}, 42:127--135, 1983.

\bibitem{OBrien:1981}
R.~W. O'Brien.
\newblock The electrical conductivity of a dilute suspension of charged
  particles.
\newblock {\em J. Colloid Interface Sci.}, 81(1):234--248, 1981.

\bibitem{OBrien:1978}
R.~W. O'Brien and L.~R. White.
\newblock Electrophoretic mobility of a spherical colloidal particle.
\newblock {\em J. Chem. Soc., Faraday Trans. II}, 74:1607--1626, 1978.

\bibitem{Ohshima:1989}
H.~Ohshima.
\newblock Approximate analytical expressions for the electrophoretic mobility
  of colloidal particles with surface-charge layers.
\newblock {\em J. Colloid Interface Sci.}, 130:281--282, 1989.

\bibitem{Overbeek:1943}
J.~Th.~G. Overbeek.
\newblock Theorie der elektrophorese.
\newblock {\em Kolloid-Beih.}, 54:287--364, 1943.

\bibitem{Prieve:1984}
D.~C. Prieve, J.~L. Anderson, J.~P. Ebel, and M.~E. Lowell.
\newblock Motion of a particle generated by chemical gradients. {Part 2}.
  electrolytes.
\newblock {\em J. Fluid Mech.}, 148:247--269, 1984.

\bibitem{Prieve:1987}
D.~C. Prieve and R.~Roman.
\newblock Diffusiophoresis of a rigid sphere through a viscous electrolyte
  solution.
\newblock {\em J. Chem. Soc. Faraday Trans. 2}, 83(8):1287--1306, 1987.

\bibitem{Saville:1979}
D.~A. Saville.
\newblock Electrical conductivity of suspensions of charged particles in ionic
  solutions.
\newblock {\em J. Colloid Interface. Sci.}, 71(3):477--490, 1979.

\bibitem{Saville:2000}
D.~A. Saville.
\newblock Electrokinetic properties of fuzzy colloidal particles.
\newblock {\em J. Colloid Interface Sci.}, 222:137--145, 2000.

\bibitem{Sten-Knudsen:2002}
O.~Sten-Knudsen.
\newblock {\em Biological Membranes: Theory of transport potentials and
  electric impulses}.
\newblock Cambridge University Press, 2002.

\bibitem{Wei:2003}
Y.~K. Wei and H.~J. Keh.
\newblock Theory of electrokinetic phenomena in fibrous porous media caused by
  gradients of electrolyte concentration.
\newblock {\em Colloids and Surfaces A: Physiochem. Eng. Aspects},
  222:301--310, 2003.

\bibitem{Wunderlich:1982}
R.~W. Wunderlich.
\newblock The effects of surface structure on the electrophoretic mobilities of
  large particles.
\newblock {\em J. Colloid Interface Sci.}, 88(2):385--397, 1982.

\end{thebibliography}
\bibliographystyle{plain}

\end{document}